\documentclass[12pt,a4paper]{article}
\usepackage[utf8]{inputenc}
\usepackage[english]{babel}
\usepackage{amsmath}
\usepackage{amsfonts}
\usepackage{amssymb}
\usepackage{graphicx}
\usepackage{lmodern}
\usepackage[left=2.5cm,right=2.5cm,top=3cm,bottom=3cm]{geometry}
\usepackage{authblk}
\usepackage[square,numbers,super,compress]{natbib}

\usepackage{rotating}

\usepackage[colorlinks,bookmarksopen,bookmarksnumbered,citecolor=red,urlcolor=red]{hyperref}
\usepackage[hyphenbreaks]{breakurl}

\usepackage{lmodern}
\usepackage[T1]{fontenc}
\usepackage{color}

\usepackage{booktabs}

\title{\vspace{-2cm}Group sequential designs for negative binomial outcomes}	

\author[a]{Tobias M{\"u}tze\thanks{Correspondence: Tobias M{\"u}tze, Institut f{\"u}r Medizinische Statistik, Humboldtallee 32, 37073 G{\"o}ttingen, Germany. Email: tobias.muetze@med.uni-goettingen.de}}
\author[b,c]{Ekkehard Glimm}
\author[b]{Heinz Schmidli}
\author[a,d]{Tim Friede}
\affil[a]{Department of Medical Statistics, University Medical Center G{\"o}ttingen, G{\"o}ttingen, Germany}
\affil[b]{Statistical Methodology, Novartis Pharma AG, Basel, Switzerland}
\affil[c]{Institute for Biometrics and Medical Informatics, Medical Faculty, Otto-von-Guericke-University Magdeburg, Magdeburg, Germany}
\affil[d]{DZHK (German Centre for Cardiovascular Research), partner site G{\"o}ttingen, G{\"o}ttingen, Germany}

\providecommand{\keywords}[1]{\textit{Keywords:} #1}
\providecommand{\refas}[1]{
\vspace{0.5cm}
\noindent
\hspace{\dimexpr-\fboxrule-\fboxsep\relax}\fbox{%
  \begin{minipage}[t]{\linewidth}
   \textit{Reference as:}\\ #1
  \end{minipage}%
}
}

\date{\vspace{-5ex}}

\usepackage{fancyhdr}
\fancypagestyle{plain}{%
  \fancyhf{}
  
  \fancyfoot[L]{\textit{Post-print}}
  \fancyfoot[C]{\thepage}
  \fancyfoot[R]{\textit{\today}}
}  
  
\begin{document}
\maketitle

\noindent\makebox[\width]{\rule{\textwidth}{0.2pt}}

\begin{abstract}
Count data and recurrent events in clinical trials, such as the number of lesions in magnetic resonance imaging in multiple sclerosis, the number of relapses in multiple sclerosis, the number of hospitalizations in heart failure, and the number of exacerbations in asthma or in chronic obstructive pulmonary disease (COPD) are often modeled by negative binomial distributions.
In this manuscript we study planning and analyzing clinical trials with group sequential designs for negative binomial outcomes.
We propose a group sequential testing procedure for negative binomial outcomes based on Wald statistics using maximum likelihood estimators.
The asymptotic distribution of the proposed group sequential tests statistics are derived.
The finite sample size properties of the proposed group sequential test for negative binomial outcomes and the methods for planning the respective clinical trials are assessed in a simulation study.
The simulation scenarios are motivated by clinical trials in chronic heart failure and relapsing multiple sclerosis, which cover a wide range of practically relevant settings.
Our research assures that the asymptotic normal theory of group sequential designs can be applied to negative binomial outcomes when the hypotheses are tested using Wald statistics and maximum likelihood estimators.
We also propose two methods, one based on Student's t-distribution and one based on resampling, to improve type I error rate control in small samples.
The statistical methods studied in this manuscript are implemented in the R package \textit{gscounts}, which is available for download on the Comprehensive R Archive Network (CRAN).
\end{abstract}

\noindent\makebox[\linewidth]{\rule{\textwidth}{0.2pt}}

\refas{
M{\"u}tze, T., Glimm, E., Schmidli, H., \& Friede, T. (2018). Group sequential designs for negative binomial outcomes. Statistical Methods in Medical Research. \burl{https://doi.org/10.1177/0962280218773115}} 

\section{Introduction}
Group sequential clinical trial designs provide a statistical framework for stopping the clinical trial early for efficacy or  futility by testing the hypothesis of interest repeatedly after groups of new observations are available.
Due to the repeated testing, hypothesis tests in group sequential designs have to be chosen such that they avoid type I error rate inflation.
Usually, the same type of test statistics are used as for the fixed sample designs, but critical values are altered to control the type I error rate.
Group sequential testing is simplified when the joint distribution of stagewise test statistics is \textit{canonical}. \cite{jennison2000group}
This is equivalent to the statement that the stagewise score statistics of the accumulating data have independent increments.
The canonical joint distribution reduces the computational effort for the calculation of the critical values and, moreover, the theory of group sequential designs is well established for test statistics following the canonical joint distribution.
However, violations of the independent increment assumption can lead to deviations of the type I error rate from the target as demonstrated by Shoben and Emerson for longitudinal data analyses based on generalized estimating equations. \cite{shoben2014violations}
For a detailed discussion of group sequential designs, we refer to the comprehensive discourses by Jennison and Turnbull, by Whitehead, and by Wassmer and Brannath. \cite{jennison2000group, whitehead1997design, wassmer2016group} \\ \indent
Counts and recurrent events are common in clinical trials.
Examples include the number of magnetic resonance imaging (MRI) lesions and relapses in relapsing multiple sclerosis, the number of exacerbations in chronic obstructive pulmonary disease (COPD) and asthma, the monthly number of migraine days following atrial septal defect closure, the number of flares in acute gouty arthritis, and the number of hospitalizations in chronic heart failure.\cite{sormani1999modelling,keene2007analysis,keene2008statistical, keene2009methods,rodes2015effect,schlesinger2011canakinumab,rogers2014analysing}
However, for group sequential designs with count and recurrent event endpoints, the literature is sparse and the methods are currently not implemented in standard software for group sequential designs such as East$^{\circledR}$ and ADDPLAN$^{\circledR}$.
The existing literature on group sequential designs with count and recurrent event data includes models such as nonparametric models, Poisson distribution, and Poisson processes with frailty. \cite{cook1996incorporating, cook1996interim, jiang1999group, xia2007procedure, cook2010sequential}
The main focus of the available literature regarding group sequential designs with recurrent events is on nonparametric and semiparametric models where the score statistics do not fulfill the independent increment structure.  \\ \indent
In this manuscript we focus on group sequential designs with  recurrent events modeled by a negative binomial distribution which is commonly considered to model endpoints in clinical trials in multiple sclerosis, chronic heart failure, asthma, and COPD. \cite{sormani1999modelling, keene2007analysis, keene2008statistical, keene2009methods, rogers2014analysing}
We propose a group sequential Wald test which follows the canonical joint distribution asymptotically.
The finite sample size performance of the test are assessed by means of a Monte Carlo simulation study with scenarios motivated by clinical trial examples.
Since the canonical joint distribution only holds asymptotically, particular emphasis is put on the performance of the proposed group sequential test for the case of small sample sizes and methods for controlling the type I error are discussed.
Moreover, we propose methods for planning clinical trials with a group sequential design and negative binomial outcomes.
The information level for negative binomial outcomes depends not only on the sample size but also on the overdispersion parameter, the event rates, and the individual follow-up times.
This provides considerable flexibility when designing a group sequential trial with negative binomial outcomes, which we discuss in detail in this manuscript. \\ \indent
The remainder of this manuscript is structured as follows.
In Section \ref{sec:Examples} we discuss the number of hospitalizations in heart failure trials and the number of MRI lesions in relapsing-remitting multiple sclerosis trials as  motivating examples for negative binomial outcomes in clinical trials.
In Section \ref{sec:groupseq} methods for analyzing group sequential designs with negative binomial outcomes are established and the respective  planning aspects are discussed in Section \ref{sec:planning}.
The operating characteristics of the methods for analyzing and planning group sequential designs with negative binomial outcomes are assessed by Monte Carlo simulation studies in Section \ref{sec:simulation}.
In Section \ref{sec:smallSample} we propose ad hoc modifications of the Wald test to achieve type I error control in the case of small sample sizes.
The examples are revisited in Section \ref{sec:examplesrev}.
A discussion of the results and of future research is provided in Section \ref{sec:discussion}.
\section{Clinical trial examples}
\label{sec:Examples}
In this section we discuss two examples of clinical trials in which the endpoint is commonly modeled by a negative binomial distribution.
Later on in this manuscript we will motivate the simulation scenarios using the examples introduced in this section.
The examples cover a wide variety of clinical trial scenarios with respect to study duration, follow-up time, and study phase.
\subsection{Chronic heart failure}
Chronic heart failure describes a medical condition in which a patient's blood flow or blood pressure is reduced.
There are three variants of this condition: heart failure with reduced ejection fraction (HFrEF), heart failure with mid-range ejection fraction (HFmrEF), and heart failure with preserved ejection fraction (HFpEF). \cite{Ponikowski2016esc}
Ejection fraction is the ratio of the amount of blood pumped out of the heart and the amount of blood in the heart chamber.
In clinical trials in heart failure the main treatment goals include the reduction of mortality and the number of heart failure hospitalizations. \cite{CHMP2016esc}
In particular when the cardiovascular mortality is low, heart failure hospitalizations are an important endpoint for characterizing the efficacy of a treatment as well as the burden for patients and the health care system.
Recently, the negative binomial distribution has been proposed as one model for heart failure hospitalizations in clinical trials in heart failure.  \cite{rogers2012eplerenone, rogers2014analysing, rogers2014effect}
In the following we focus on heart failure hospitalizations in clinical trials in HFpEF.
For more information on HFpEF we refer to the literature. \cite{udelson2011, redfield2016}
The CHARM-Preserved trial is a double-blind, multi-center, placebo-controlled phase 3 clinical trial which assessed the efficacy of candesartan in patients with HFpEF. \cite{yusuf2003effects}
A total of 3023 patients were randomly assigned to either candesartan (1514 patients) or placebo (1509 patients); the median follow-up time was 36.6 months.
The heart failure hospitalizations observed during the CHARM-Preserved trial are listed in Table \ref{table:hfhCharmed}.
\begin{table}[ht]
\caption{Heart failure hospitalizations in the CHARM-Preserved trial. \cite{yusuf2003effects}}
\label{table:hfhCharmed}
\begin{center}
\begin{tabular}{l*{1} {c}{c}}
\toprule
& \textbf{Placebo} &  \textbf{Candesartan} \\ \midrule
Number of patients & 1509 & 1514 \\
Total follow-up years & 4374.03 & 4424.62\\
Patients with $\geq 1$ admission & 278 & 230\\
Total admissions & 547 & 392\\
\bottomrule
\end{tabular}
\end{center}
\end{table}\\
Using the negative binomial distribution, the rate ratio for recurrent heart failure hospitalizations is estimated to be $\hat{\theta}=0.71$. \cite{rogers2014analysing}\\ \indent
In general, adaptive trial designs have been recommended as one potential design for increasing the effectiveness of clinical trials for cardiovascular diseases. \cite{mehta2009optimizing,jackson2015improving}
\subsection{Relapsing-remitting multiple sclerosis}
Relapsing-remitting multiple sclerosis is a form of multiple sclerosis which is characterized by sudden worsenings of the symptoms, relapses, followed by a remission period with no or only few symptoms.
The therapeutic goal for relapsing-remitting multiple sclerosis is the prevention of disease progression, i.e., the prevention of relapses and worsening of disabilities.
However, in particular in early phases, the number of lesions, measured through magnetic resonance imaging (MRI), is chosen as an endpoint instead of the number of relapses.
This is due to the number of MRI lesions generally being more sensitive to disease activity and progression which is associated with a smaller required sample size in clinical trials.
More precisely, the disease activity is often measured through the number of new gadolinium-enhancing lesions on T1-weighted images or through the number of new or enlarged lesions on T2-weighted images.
It is also possible to combine both types of lesions which is then referred to as the number of combined unique active MRI lesions (CUALs).
For a broader discussion of this topic, see Cohen and Rudick. \cite{cohen2007multiple}
Selmaj \textit{et al.} reported an adaptive, dose-ranging, randomized phase 2 study assessing the dose-response relationship of siponimod (the BOLD study). \cite{selmaj2013siponimod}
The primary endpoint was the dose-response relationship assessed considering the number of CUALs.
Table \ref{table:msSipo} lists the monthly number of CUALs at 3 months for placebo and two of the five doses from the BOLD study.
The dosage of 0.25 mg per day was the smallest dose and 2 mg per day the dosage after which only a small gain in efficacy was observed.
\begin{table}[ht]
\begin{center}
\caption{Monthly number of CUALs (at 3 months) reported in the BOLD study. \cite{selmaj2013siponimod}}
\label{table:msSipo}
\begin{tabular}{l*{1} {c}{c}{c}}
\toprule
& \textbf{Placebo} &  \textbf{Siponimod 0.25 mg} &   \textbf{Siponimod 2 mg} \\ \midrule
Number of patients & 61 & 51 & 45\\
Monthly CUALs & 1.39 & 0.78 & 0.42\\
Lesion ratio to placebo & - & 0.557 & 0.303\\
\bottomrule
\end{tabular}
\end{center}
\end{table}\\
The number of combined unique active lesions was analyzed using a negative binomial model.
Moreover, it is worthwhile to mention that the study consisted of two patient cohorts of which cohort one was considered for an interim analysis and for selecting additional doses for the second cohort.
\section{Analyzing group sequential designs with negative binomial outcomes}
\label{sec:groupseq}
In this section we introduce the statistical model and propose a Wald test for analyzing group sequential designs with negative binomial outcomes.
\subsection{Statistical model, hypotheses, and notation}
We define $K$ as the maximum number of \textit{data looks} or \textit{analyses}, that is the maximum number of times the hypotheses of interest can be tested throughout the conduct of the clinical trial.
For instance, a design with $K=3$ data looks has two interim analyses and one final analysis if the trial is not stopped early.
We model the number of events of patient $j=1,\ldots, n_{i}$ in treatment group $i=1,2$ at data look $k=1,\ldots, K$ as a homogeneous Poisson point process with rate $\lambda_{ij}$.
Let $t_{ijk}$ be the exposure time of patient $j$ in treatment group $i$ at look $k$.
After an exposure time of $t_{ijk}$, the number of events $Y_{ijk}$ of patient $j$ in treatment group $i$ at data look $k$ given the rate $\lambda_{ij}$ is Poisson distributed with rate $t_{ijk}\lambda_{ij}$, that is
\begin{align*}
Y_{ijk}|\lambda_{ij} \sim \operatorname{Pois}(t_{ijk} \lambda_{ij}).
\end{align*}
The between patient heterogeneity of the event rates is modeled by assuming that the subject specific event rates $\lambda_{ij}$ are Gamma distributed.
The Gamma distribution of $\lambda_{ij}$ has a shape $\alpha=1/\phi$ and rate $\beta=1/(\phi \mu_i)$, i.e.,
\begin{align*}
\lambda_{ij} \sim \Gamma\left(\frac{1}{\phi}, \frac{1}{\phi \mu_i}\right).
\end{align*}
Then, the random variable $Y_{ijk}$ is marginally negative binomial distributed with rate $t_{ijk}\mu_i$ and shape parameter $\phi$, i.e.,
\begin{align*}
Y_{ijk} \sim \operatorname{NB}\left(t_{ijk}\mu_i, \phi\right).
\end{align*}
For $y\in \mathbb{N}_{0}$, the probability mass function of $Y_{ijk}$ is given by
\begin{align*}
\mathbb{P}\left(Y_{ijk}=y\right)=\frac{\Gamma(y + 1/\phi)}{\Gamma(1/\phi)y!}\left(\frac{1}{1+\phi t_{ijk} \mu_i}\right)^{1/\phi} \left(\frac{\phi t_{ijk}\mu_i}{1 + \phi t_{ijk}\mu_i }\right)^{y}.
\end{align*}
The expected value and the variance of the random variable $Y_{ijk}$ are given by
\begin{align*}
\mathbb{E}\left[Y_{ijk}\right] = t_{ijk} \mu_i\quad \text{ and } \quad
\operatorname{Var}\left[Y_{ijk}\right] = t_{ijk} \mu_i (1 + \phi t_{ijk} \mu_i).
\end{align*}
As the shape parameter approaches zero, the negative binomial distribution converges to a Poisson distribution.
While a Poisson process has independent, Poisson distributed increments,\cite{cook2007statistical}
a negative binomial counting process only has conditionally independent increments, i.e., $(Y_{ijk} - Y_{ij(k-1)})|\lambda_{ij}$ $(k=2,\ldots, K)$ are Poisson distributed with rate $(t_{ijk} - t_{ij(k-1)})\lambda_{ij}$.
Marginally, the increments are negative binomial distributed and dependent with the covariance
\begin{align*}
\operatorname{Cov}\left[Y_{ij(k-1)}, Y_{ijk} - Y_{ij(k-1)}\right]
= t_{ijk}(t_{ijk} - t_{ij(k-1)})\phi \mu_{i}^{2}.
\end{align*}
\indent In this manuscript we consider smaller values of $\mu_i$ to be better.
The statistical hypothesis testing problem of interest is
\begin{align*}
H_0: \frac{\mu_1}{\mu_2}\geq \delta \quad \text{vs.} \quad H_1: \frac{\mu_1}{\mu_2} < \delta.
\end{align*}
For $\delta\in (1,\infty)$ non-inferiority of treatment 1 compared to treatment 2 is tested and for $\delta \in (0,1]$ superiority of treatment 1 compared to treatment 2 is tested.
We denote the rate ratio as $\theta=\mu_1/\mu_2$.
Let the log-rate be $\beta_{i} = \log(\mu_i)$.
\subsection{Maximum likelihood estimators and Wald statistics}
In the following we discuss the Wald test for the hypothesis $H_0$ at data look $k=1,\ldots, K$.
Let $\hat{\beta}_{ik}$ be the maximum likelihood estimator of the log-rate $\beta_i$ and $\hat{\phi}_{k}$ be the maximum likelihood estimator of the shape parameter $\phi$ based on the data available at data look $k$.
Then, the maximum likelihood estimator for the rate $\mu_i$ at data look $k$ is given by $\hat{\mu}_{ik}=\exp(\hat{\beta}_{ik})$.
The maximum likelihood estimator $(\hat{\beta}_{1k}, \hat{\beta}_{2k}, \hat{\phi}_{k})$ at data look $k$ is obtained by maximizing the likelihood function
\begin{align*}
L(\beta_1, \beta_2, \phi|Y_{ijk} = y_{ijk}, j = 1, \ldots, n_i, i = 1,2)
= \prod\limits_{i=1}^{2}\prod\limits_{j=1}^{n_i} \mathbb{P}\left(Y_{ijk}=y_{ijk}\right).
\end{align*}
The counts observed at previous looks do not have to be considered in the likelihood since the random variables $Y_{ijk}$ $(j = 1, \ldots, n_i, i = 1,2)$ are sufficient for the parameter vector $(\beta_1, \beta_2, \phi)$ at data look $k$.
As a reviewer pointed out, it follows from basic properties of the Poisson distribution that the distribution of $Y_{ij1}, \ldots, Y_{ij(k-1)} | Y_{ijk}, \lambda_{ij}$ does not depend on the subject specific event rate $\lambda_{ij}$  and, consequently, the marginal distribution $Y_{ij1}, \ldots, Y_{ij(k-1)} | Y_{ijk}$ also does not depend on $\lambda_{ij}$ either.
Details about the sufficiency of $Y_{ijk}$ at data look $k$ and the calculation of the maximum likelihood estimators are presented in Appendix \ref{sec:ApendA}.
It should be noted that no closed form expression is available for the maximum likelihood estimators with negative binomially distributed endpoints.
Moreover, the maximum likelihood rate estimator is not identical to the method of moments rate estimator which estimates a rate by the ratio of total events and total exposure time.
They are only identical if the exposure times within a treatment arm are identical.
The Fisher information of the parameter $\beta_i$ at data look $k$ is given by
\begin{align*}
I_{\beta_{i}}^{(k)} =
\sum_{j=1}^{n_i} \frac{ t_{ijk} \exp(\beta_{i})}{1+\phi t_{ijk} \exp(\beta_{i}) } =
\sum_{j=1}^{n_i} \frac{ t_{ijk} \mu_i}{1+\phi t_{ijk} \mu_i }.
\end{align*}
The expected Fisher information is a special case of the  expected Fisher information of the log-rates in a negative binomial regression model. \cite{lawless1987negative}
The larger the Fisher information, the smaller the uncertainty and the larger the knowledge about the parameter $\beta_{i}$.
The Fisher information increases as the shape parameter decreases, the sample size $n_i$ increases, or the exposure time $t_{ijk}$ increases. \\ \indent
The Wald statistic for testing the hypothesis $H_{0}$ at data look $k=1,\ldots, K$ is based on the difference of log-rates, i.e., $\hat{\beta}_{1k} - \hat{\beta}_{2k}$.
Let the information level at data look $k$ be
\begin{align}
\label{eq:infoK}
\mathcal{I}_k = \frac{1}{\frac{1}{I_{\beta_{1}}^{(k)}} + \frac{1}{I_{\beta_{2}}^{(k)}}}
= \frac{I_{\beta_{1}}^{(k)}I_{\beta_{2}}^{(k)}}{I_{\beta_{1}}^{(k)} + I_{\beta_{2}}^{(k)}}.
\end{align}
Analogously to the Fisher information, the information level measures the knowledge available about the difference of the log-rates.
The effect of shape parameter, sample size, and exposure time on the information level are as described above.
The differences of log-rates is asymptotically normally distributed, i.e.,
\begin{align*}
  \left(\hat{\beta}_{1k} - \hat{\beta}_{2k} - \left(\beta_1 - \beta_2\right)\right)\sqrt{\mathcal{I}_k}
  \stackrel[n\to \infty]{\mathcal{D}}{\sim} \mathcal{N}\left(0, 1\right).
\end{align*}
The asymptotic normality of the differences of log-rates follows from the asymptotic normality of the vector $(\hat{\beta}_{1k}, \hat{\beta}_{2k}, \hat{\phi})$ shown by Lawless (1987).\cite{lawless1987negative}
The information level $\mathcal{I}_k$ depends on the shape parameter $\phi$ as well as the true, unknown rates.
To construct a test statistic for $H_0$ using the asymptotic normality of the difference of log-rates, we replace the information level $\mathcal{I}_{k}$ by a consistent estimator $\hat{\mathcal{I}}_{k}$ and obtain
the Wald statistic at data look $k$ as
\begin{align*}
  T_{k}= \left(\hat{\beta}_{1k} - \hat{\beta}_{2k} - \log(\delta)\right) \sqrt{\hat{\mathcal{I}}_{k}}.
\end{align*}
Since we estimate the information level by a consistent estimator, the Wald statistic $T_{k}$ is still asymptotically normally distributed under the null hypothesis due to Slutsky's theorem.
Under the alternative hypothesis, the Wald statistic $T_{k}$ is approximately normally distribution, i.e.,
\begin{align}
  \label{eq:distTeststatH1}
  T_{k} \sim \mathcal{N}\left(\beta_1 - \beta_2 - \log(\delta), 1 \right) = \mathcal{N}\left(\log(\theta/\delta), 1 \right).
\end{align}
A consistent estimator $\hat{\mathcal{I}}_{k}$ for the information level $\mathcal{I}_{k}$ is obtained by plugging in the maximum likelihood rate and shape estimators into \eqref{eq:infoK}.
The hypothesis $H_0$ is rejected if $T_k$ is smaller than or equal to the critical value $c_k$ for any $k=1,\ldots, K$.
In the next subsection we discuss the calculation of the critical values $c_k$.
\subsection{Group sequential testing}
In the following we recapitulate the general idea of group sequential testing and discuss how the canonical joint distribution applies to the Wald test defined above.
We focus on calculating the critical values for group sequential designs which allow stopping for efficacy.
In practice it is very common to also include futility stopping boundaries. However, these are almost always treated as non-binding such that they do not affect the efficacy stopping boundaries. \cite{gallo2014alternative}
A design with stopping for efficacy is discontinued at data look $k=1,\ldots, K-1$ if the null hypothesis $H_0$ can be rejected.
If the null hypothesis $H_0$ cannot be rejected at data look $k=1,\ldots, K-1$, the study is continued.
While the efficacy stopping boundaries are calculated in this way, it is advisable to also carefully consider non-binding futility stopping rules, in particular with respect to their impact on operating characteristics such as power, sample size, trial duration etc.
Furthermore, a failure to specify futility rules during the planning stage might result in the usage of inappropriate ad-hoc futility rules with undesired or suboptimal operation characteristics.
Therefore, we examine non-binding futility boundaries in Section \ref{sec:planning} when discussing planning a group sequential design with negative binomial outcomes.\\ \indent
A group sequential test for efficacy controls the type I error rate $\alpha$ if the critical values $c_{1}, \ldots, c_{K}$ are chosen such that the probability to incorrectly reject the null hypothesis at one of the data looks is equal to $\alpha$:
\begin{align*}
\alpha = \mathbb{P}_{H_0}\left(T_{k} \leq c_{k} \quad \text{for some } k=1,\ldots, K\right).
\end{align*}
To obtain a solution, the global type I error rate $\alpha$ is split into $K$ local type I error rates $\pi_{k}$, $k=1,\ldots,K$, such that the local type I error rates sum up to $\alpha$, i.e., $\sum_{k}\pi_{k}=\alpha$.
Then, the critical values $c_{k}$, $k=1,\ldots,K$, are successively calculated such that
\begin{align}
\label{eq:criticCal}
\mathbb{P}_{H_0}\left(T_{1} > c_{1},\ldots, T_{k-1} > c_{k-1},  T_{k} \leq c_{k}\right) = \pi_k.
\end{align}
For calculating the critical value $c_{k}$ the joint distribution of test statistics $T_{1},\ldots, T_{k}$ under the null hypothesis $H_0$ must be known.
The vector of Wald statistics $(T_{1},\ldots, T_{k})$ follows at the boundary $\partial H_{0}$ asymptotically a multivariate normal distribution with mean vector $0\in \mathbb{R}^{k}$ and $k\times k$-dimensional covariance matrix $\Sigma_{k}$ which $(k_{1}, k_{2})$-th and $(k_{2}, k_{1})$-th  entries are given by
\begin{align}
\label{eq:jointdistCov}
\sqrt{\frac{\mathcal{I}_{k_{1}}}{\mathcal{I}_{k_{2}}}}, \qquad 1\leq k_{1} \leq k_{2} \leq K.
\end{align}
Here, $\mathcal{I}_{k}$ is the information level at look $k$.
This type of distribution is often referred to as the \textit{canonical joint distribution}. \cite{jennison2000group}
The asymptotic joint distribution of the sequence of test statistics follows from the more general result that the canonical joint distribution is the limiting distribution of a sequence of semiparametric efficient test statistics in a semiparametric model. \cite{scharfstein1997semiparametric}
This class of models and test statistics includes parametric models and Wald statistics, see the remark on page 1344 in Scharfstein \textit{et al.} (1997). \cite{scharfstein1997semiparametric}
The negative binomial model introduced above fulfills this requirement since the random variables $Y_{ijk}$ with $j=1,\ldots, n_i$ and $i=1,2$ are sufficient for the parameter vector $(\beta_1, \beta_2, \phi)$ at data look $k$.
Thus, the sequence of Wald statistics in the negative binomial model converges in distribution to the canonical joint distribution.
It is worth highlighting that the asymptotic joint distribution of the sequence of Wald statistics in the negative binomial model also follows from arguments made by Jennison and Turnbull (1997) in their discussion.\cite{jennison1997group}
The asymptotic joint distribution is obtained even though the number of events of a patient at disjunct time periods is dependent.
It is important to emphasize that estimating the log rates in the Wald statistics with the maximum likelihood estimators and not with the method of moments estimators is mandatory to obtain asymptotically the canonical joint distribution.
With the sequence of Wald statistics asymptotically following the joint canonical distribution, the critical value $c_{k}$ is calculated by solving \eqref{eq:criticCal} under the assumption of asymptotic normality.
For $k=1$, the critical value is the $\pi_1$-quantile of a standard normal distribution, i.e., $c_1 = q_{\pi_1}$.\\ \indent
It remains to discuss how to determine the data look specific type I error rates $\pi_{k}$.
The most common method for allocating the global type I error rate $\alpha$ is the \textit{error spending approach}.\cite{lan1983discrete}
An error spending function is a non-decreasing function $f:[0,\infty)\to [0,\alpha]$ with $f(0)=0$ and $f(t)=\alpha$  for $t\geq 1$.
Then, the type I error rate $\pi_k$ allocated to data look $k=1,2,\ldots, K$ is defined by
\begin{align*}
\pi_1 &= f\left(\mathcal{I}_{1} / \mathcal{I}_{max}\right),\\
\pi_k &= f\left(\mathcal{I}_k / \mathcal{I}_{max}\right) - f\left(\mathcal{I}_{k-1} / \mathcal{I}_{max}\right)\quad k=2,3,\ldots, K.
\end{align*}
Here, $\mathcal{I}_{max}$ denotes the prespecified maximum information level at which the final data look is performed if the trial is not stopped early for efficacy.
\subsection{Determining the critical values}
\label{sec:critical}
To calculate the critical value $c_{k}$, the canonical joint distribution of the Wald statistics and the allocated type I error rate $\pi_{k}$ must be known.
Both depend on the information levels $\mathcal{I}_{1}, \ldots, \mathcal{I}_{k}$.
When the data looks are performed at information levels specified prior to the trial, the critical values can be determined prior to the trial, too.
Then, the main issue is to monitor the information level of the clinical trial.
Here, we focus on the scenario where the information levels of the data looks are not specified before the trial and can differ from the pre-planned information levels.
We assume that future information levels are conditionally independent of previous treatment effect estimates given previous information levels.
Violations of this assumption can result in an inflation of the type I error rate.\cite{proschan1992effects}
If the information level at a data look is unknown, it has to be estimated and the respective critical value is calculated based on the estimated information level.
The information levels are functions of the exposure times, the rates, and the shape parameter.
While the exposure times at look $k$ become known, the rates and the shape parameter remain unknown and must be estimated from the available data.
Let $\hat{\mathcal{I}}_{k}$ be the plug-in estimator of the information level $\mathcal{I}_k$ at data look $k$ obtained through the maximum likelihood estimators $(\hat{\beta}_{1k}, \hat{\beta}_{2k}, \hat{\phi}_k)$.
Then, the type I error rate $\pi_k$ allocated to data look $k$ is given by
\begin{align*}
\pi_1 &= f\left(\hat{\mathcal{I}}_{1} / \mathcal{I}_{max}\right),\\
\pi_k &= f\left(\hat{\mathcal{I}}_k / \mathcal{I}_{max}\right) - f\left(\hat{\mathcal{I}}_{k-1} / \mathcal{I}_{max}\right),\quad k=2,3,\ldots, K-1.
\end{align*}
It is important to emphasize that the information level estimator $\hat{\mathcal{I}}_{k-1}$ calculated at the previous data look is not updated at data look $k$.
When the theoretical information gain between looks $k-1$ and $k$ is small, it might occur that the estimated information decreases, that is $\hat{\mathcal{I}}_{k}<\hat{\mathcal{I}}_{k-1}$.
In this case, no type I error rate is allocated to data look $k$, i.e., $\pi_k = 0$ and the null hypothesis cannot be rejected at data look $k$.
Moreover, at the end of the trial, i.e., at data look $K$, the estimated information level $\hat{\mathcal{I}}_{K}$ usually differs from the anticipated maximum information level $\mathcal{I}_{max}$.
Therefore, the remaining type I error rate is allocated to data look $K$, i.e.,
\begin{align*}
\pi_{K} = \alpha - f\left(\hat{\mathcal{I}}_{K-1} / \mathcal{I}_{max}\right).
\end{align*}
When the type I error rate $\alpha$ is allocated based on information estimates, the allocated rates still sum up to $\alpha$.
In addition to $\pi_{k}$, the covariance matrix $\Sigma_{k}$ of the canonical joint distribution has to be estimated to calculate the critical value $c_{k}$.
The entries of the covariance matrix \eqref{eq:jointdistCov} are estimated by plugging in the respective information level estimators, with the estimator for the information level at data look $k$ only using the data available up to data look $k$. \\ \indent
The maximum likelihood estimators of the rates and the shape parameter are consistent estimators.
Since the Fisher information, the information levels, the information level ratios, and the allocation of the type I error rate are continuous functions in the rates and the shape parameter, the respective plug-in estimators are consistent estimators, too.
Thus, the discussed asymptotic considerations about the Wald test in group sequential designs with negative binomial outcomes still hold.
\section{Planning aspects of group sequential designs}
\label{sec:planning}
An important part of planning a clinical trial is to determine the sample size.
An adequately planned sample size assures that the trial is able to detect a relevant treatment effect with high probability.
In group sequential designs the sample size becomes a random variable and the maximum information and the (information) time of the data looks are fixed.
Highlighting the specifics for the case of a negative binomial outcomes, in the following we recapitulate the general approach for power and sample size planning in group sequential designs. \cite{jennison2000group}
We start with group sequential designs which only include stopping for efficacy and conclude with designs additionally including non-binding futility boundaries.
Under the alternative $H_1$, that is for $\theta<\delta$, the power of a group sequential design for a given maximum number of looks $K$ and a set of critical values $c_1, \ldots, c_{K}$ is given by
\begin{align}
\label{eq:power}
\operatorname{Power} = 1 - \mathbb{P}_{\theta}\left(T_{1} > c_{1}, \ldots, T_{K} > c_{K}\right).
\end{align}
As stated in \eqref{eq:distTeststatH1}, under the alternative $H_1$, the test statistic $T_k$ can be approximated by a normal distribution. Moreover, the vector of test statistics  $(T_1, \ldots, T_K)$ can be approximated by a multivariate normal distribution with mean vector
\begin{align*}
\boldsymbol{\mu} = \left(\sqrt{\mathcal{I}_{1}}\left(\log\left(\theta\right)-\log(\delta)\right), \ldots, \sqrt{\mathcal{I}_{K}} \left( \log\left(\theta\right)-\log(\delta)\right) \right)^{\prime}
\end{align*}
and covariance matrix as defined in \eqref{eq:jointdistCov}.
For calculating the power and the sample size, the parameter under the alternative hypothesis $H_1$ is fixed at $\theta = \theta^{*}$.
However, the information levels $\mathcal{I}_k$ $(k=1,\ldots, K)$ of the data looks are unknown.
In order to decide about the placement of the data looks, it is convenient to express the information level $\mathcal{I}_{k}$ as a fraction of the maximum information $\mathcal{I}_{max}$, i.e.,
$\mathcal{I}_{k} = w_k \mathcal{I}_{max}, k = 1,\ldots, K, w_k \in (0,1]$.
The ratio $w_{k}$ is prespecified and denotes the information fraction at which the $k$-\textit{th} data look is conducted.
Then, the mean vector and the covariance of the canonical joint distribution can be simplified to only depend on the maximum information level, i.e., the mean vector is given by
\begin{align*}
\boldsymbol{\mu} = \left(\log(\theta^{*})-\log(\delta)\right) \sqrt{\mathcal{I}_{max}} \left(\sqrt{w_{1}}, \ldots, \sqrt{w_{K}} \right)^{\prime}
\end{align*}
and the $(k_1, k_2)$-\textit{th} and $(k_2, k_1)$-\textit{th} entry of the covariance matrix reduce to $\sqrt{w_{k_1}/w_{k_2}}$.
The maximum information level $\mathcal{I}_{max}$ is now the only missing variable in the power function and the maximum information $\mathcal{I}^{(1-\beta)}_{max}$ required to obtain a prespecified power can be calculated by solving \eqref{eq:power}. 
It is worth emphasizing that achieving the target power can only be guaranteed if the data looks are performed when the observed information level ratios reach the prespecified ratios $w_{k},\, k=1,\ldots, K$. If the information level ratios of the data looks are changed during the conduct of the clinical trial while the maximum information $\mathcal{I}^{(1-\beta)}_{max}$ remains unchanged, the actual power might differ from the target power. Moreover, performing the data looks at prespecified information times requires that the information is monitored. We revisit the issue of information monitoring in the discussion.
As of now, we discussed calculating the maximum information $\mathcal{I}_{\max}^{(1-\beta)}$ required to obtain a prespecified power in a clinical trial with a group sequential design and negative binomial outcomes.
Next, we focus on the actual sample size.
The sample sizes $n_{1}$ and $n_2$ required to obtain the power $1-\beta$ can be calculated using the maximum information level $\mathcal{I}_{\max}^{(1-\beta)}$.
Let $\kappa = n_2/n_1$ be the randomization ratio, $t_{ij}=t_{ijK}$ the exposure times at the end of the trial, and $\mu_{1}^{*}$, $\mu_{2}^{*}$ the rates under the alternative hypothesis $H_1$.
Then, the sample size $n_1$ required to obtain a power of $1-\beta$ is obtained by solving
\begin{align*}
\frac{\left(\sum_{j=1}^{n_1}\frac{t_{1j}\mu^{*}_{1}}{1+\phi t_{1j} \mu_{1}^{*}}\right)\left(\sum_{j=1}^{\kappa n_1}\frac{t_{2j}\mu^{*}_{2}}{1+\phi t_{2j} \mu_{2}^{*}}\right)}{\sum_{j=1}^{n_1}\frac{t_{1j}\mu^{*}_{1}}{1+\phi t_{1j} \mu_{1}^{*}} + \sum_{j=1}^{\kappa n_1}\frac{t_{2j}\mu^{*}_{2}}{1+\phi t_{2j} \mu_{2}^{*}}} = \mathcal{I}^{(1-\beta)}_{max}.
\end{align*}
This equation highlights that the sample size $n_1$ can only be calculated with assumptions on the exposure times $t_{ij}$.
Common approaches include planning the study with a minimum or in general an identical exposure time, i.e., $t_{ij}=t$.
In the latter case the sample size $n_1$ is given by
\begin{align*}
n_{1} = \mathcal{I}^{(1-\beta)}_{max} \left(\frac{\mu_1 + \kappa \mu_2}{t \kappa \mu_1 \mu_2} + (1+\kappa) \phi\right).
\end{align*}
\indent We conclude this section with a discussion of the planning of a group sequential design which includes non-binding futility boundaries in addition to the efficacy boundaries.
Non-binding futility rules for group sequential designs are calculated such that the type I error rate is still maintained if the non-binding futility rules are not applied.
However, non-binding futility rules can affect the power.
In the following, we briefly outline the calculation of the maximum information and non-binding futility boundaries such that the power is achieved based on the error spending approach, see also Section 7 in Jennison and Turnbull (2000). \cite{jennison2000group}
In a group sequential design with non-binding futility, the efficacy boundaries are calculated as in a group sequential design without futility stopping which we discussed above.
Let $c_k$ be the efficacy boundary and $\pi_k$ the allocated type I error rate at data look $k=1,\ldots, K$.
The power $1-\beta$ is set at the alternative $\theta=\theta^{*}$.
For planning purposes we prespecify the information fraction $w_k = \mathcal{I}_k / \mathcal{I}_{max}$ at data look $k = 1,\ldots, K$.
Let $g:[0,\infty)\to [0,\beta]$ with $g(0)=0$ and $g(t)=\beta$ for $t\geq 1$ be the type II error spending function.
Here, $\beta$ denotes the target type II error rate, i.e., the target power is $P=1-\beta$.
To determine the non-binding futility boundaries, the type II error rate $\beta$ is allocated to the different data looks using the type II error spending function:
\begin{align*}
  \zeta_1 & = g\left(\mathcal{I}_1 / \mathcal{I}_{\max}\right), \\
  \zeta_k & = g\left(\mathcal{I}_k / \mathcal{I}_{\max}\right) - g\left(\mathcal{I}_{k-1} / \mathcal{I}_{\max}\right), \quad k=2,\ldots, K.
\end{align*}
Here, $\zeta_k$ denotes the type II error rate allocated to data look $k$.
Then, the non-binding futility boundaries $d_k$ $(k=1,\ldots, K)$ and the maximum information level $\mathcal{I}_{max}$ to obtain the type II error rate $\beta$  are determined by solving the following system of equations numerically:
\begin{align*}
\zeta_{k} & = \mathbb{P}_{\theta^{*}}\left(c_{1} < T_{1} < d_{1}, \ldots, T_{k} \geq d_{k}\right), \quad k = 1,\ldots, K,\\
c_K & = d_K.
\end{align*}
%
%
\section{Simulation study}
\label{sec:simulation}
The theoretical considerations of group sequential designs with negative binomial outcomes rely on the asymptotic normality of the Wald statistics and that the joint distribution of the Wald statistics from different data looks asymptotically follows  the canonical joint distribution.
Therefore, in this section we assess the operating characteristics of the proposed group sequential design for negative binomial distributed outcomes for finite sample sizes by means of Monte Carlo simulation studies.
The operating characteristics are the type I error rate, the power, the expected trial duration, the expected sample size, and the expected information level.
Particular emphasis is put on the potential benefits of group sequential designs compared to fixed sample designs for negative binomial outcomes.
The type I error rate is assessed to study whether a group sequential design inflates the type I error rate compared to a fixed sample design.
Such an assessment is necessary for asymptotic tests because the sample size at the interim analyses is in general lower compared to the fixed sample design which can increase the type I error rate of a test in the group sequential design compared to the fixed sample design.
The choice of simulation scenarios is motivated by the clinical trial examples discussed in Section \ref{sec:Examples}.
The clinical trial examples result in two very different simulation setups.
The annualized rates in the scenarios motivated by the number of CUALs in relapsing-remitting multiple sclerosis is about tenfold the annualized hospitalization rates in chronic heart failure with preserved ejection fraction.
The sample size in the clinical trial in multiple sclerosis is substantially smaller than the sample size in the heart failure trial (around 100 versus around 1000 patients per arm).
In the scenarios motivated by the multiple sclerosis trial a fixed exposure time of 6 months at the end of study is assumed.
In the scenarios motivated by the heart failure trial patients are uniformly recruited over a period of 15 months and followed until the study ends after 48 months, i.e., the exposure times vary between 33 and 48 months.
For the simulations, we select the patients' recruitment times in a deterministic manner.
The shape parameter is similar in the two motivating examples with $\phi=2,3,4$ for the multiple sclerosis scenarios and $\phi=2,3,4, 5$ for the chronic heart failure scenarios.
It is important to note that even though the shape parameters are similar in the two sets of scenarios, the variances and the overdispersion of the counts are very different.
For negative binomial outcomes the variance is $t\mu(1+\phi t\mu)$ and the index of overdispersion $D$, that is the ratio of the variance and the expected value, is equal to $D=1+t\mu \phi$.
While the shape parameter is similar in the two sets of scenarios, the rates are not similar.
Therefore, the variance and the overdispersion in the scenarios for clinical trials in multiple sclerosis are much larger than in the scenarios for clinical trials in chronic heart failure.
The simulation study is conducted for a test of superiority, i.e. $H_{0}: \mu_1 / \mu_2 \geq \delta=1$.
The one-sided significance level is $\alpha=0.025$ and the target power under the alternative $\theta=\theta^{*}$ is $1-\beta = 0.8, 0.9$.
The parameters for the simulation study are listed in Table \ref{table:scenarios}.
\begin{table}[h]
\begin{center}
\caption{Scenarios considered in the simulation study motivated by the clinical trial examples from Section \ref{sec:Examples}.}
\label{table:scenarios}
\begin{tabular}{l*{1} {c}{c}{c}{c}}
\toprule
 &  \textbf{Multiple sclerosis} &  \textbf{Heart failure} \\ \midrule
 One-sided significance level $\alpha$ & 0.025 & 0.025\\
 Superiority margin $\delta$  & 1 & 1\\
Shape parameter $\phi$ & 2, 3, 4 & 2, 3, 4, 5\\
Data looks $K$ & 2, 3 & 2, 3, 5\\
Individual follow-up  [years] & 0.5  & 2.75--4 \\
Recruitment period [years] & 1.5  & 1.25 \\
Study duration [years] & 2 & 4  \\ \midrule
\textbf{Simulations under $H_0$}\\
Maximum sample sizes $n_1=n_2$ & $50, 80, \ldots, 230$ & $800, 1100, 1400$\\
Annualized rates $\mu_1=\mu_2$ & 6, 8, 10 & 0.08, 0.1, 0.12, 0.14\\
\textbf{Simulations under $H_1$}\\
Annualized rate $\mu_2$ & 8.4 & 0.125\\
Rate ratio $\theta^{*}=\mu_1 / \mu_2$ & 0.5, 0.7 & 0.7, 0.8\\
Target power $1-\beta$ & 0.8, 0.9 & 0.8, 0.9\\
\bottomrule
\end{tabular}
\end{center}
\end{table}\\
The data looks are performed at prespecified calendar times at which an information level of $\mathcal{I}_{k} = k\mathcal{I}_{max}/K$ $(k=1,\ldots, K)$ is expected based on the parameters of the respective scenario.
The observed information level $\hat{\mathcal{I}}_{k}$ at which the data look is performed might differ from the preplanned information level $\mathcal{I}_{k}$.
The type I error rate allocation and the critical value calculation at each data look $k=1,
\ldots, K$ are performed based on the observed information level $\hat{\mathcal{I}}_{k}$.
Performing the simulations with data looks at prespecified calendar times reduces the computational effort because no information monitoring is required.
Moreover, since the type I error is allocated and the critical values are calculated based on the observed information level, any potential effect due to estimating the information levels is still incorporated.
In the simulations, the shape parameter $\phi$ is estimated as it would be in practical applications. \\ \indent
We consider group sequential designs with the error spending functions which give critical values similar to the ones for Pocock's test and O'Brien \& Fleming's test,\cite{lan1983discrete} respectively, i.e.,
\begin{align*}
f_{P}(x) &= \min \left\{\alpha \log\left(1+(e-1)x\right), \alpha \right\},\\
f_{OF}(x) &= \min \left\{2\left(1-\Phi\left(\frac{q_{\alpha/2}}{\sqrt{x}}\right)\right), \alpha \right\}.
\end{align*}
Here, $e=\exp(1)$ denotes Euler's number and $q_{\alpha}$ denotes the $\alpha$-quantile of a standard normal distribution.
We present the simulation results for the scenarios motivated by the number of heart failure hospitalizations and the results for the scenarios motivated by the number of lesions separately, starting with the results for the heart failure hospitalization scenarios.
\subsection{Heart failure hospitalizations}
\label{sec:simHF}
\begin{figure}[ht]
\centering
\includegraphics[width=1\textwidth]{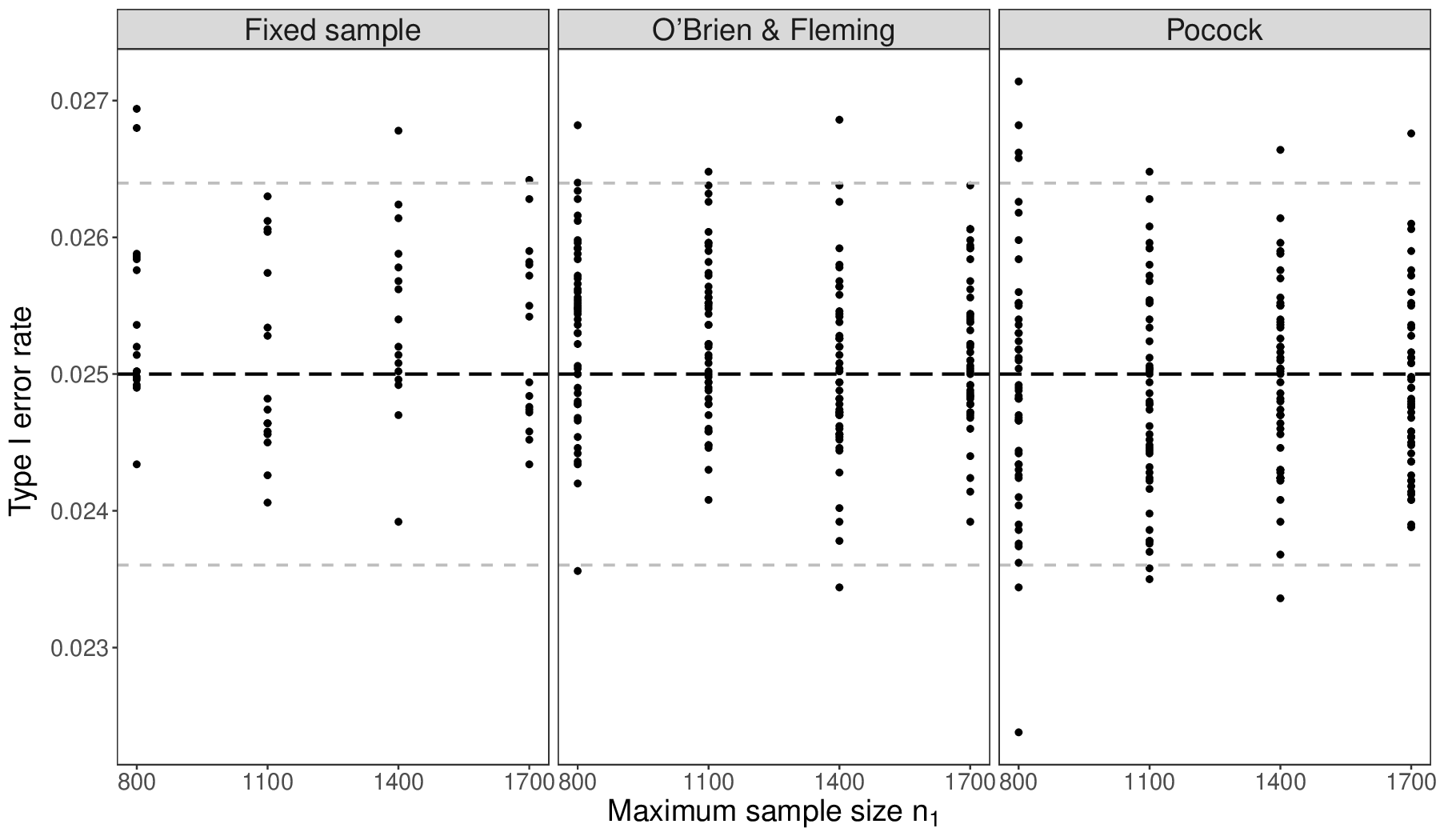}
\caption{Type I error rate of the Wald test for negative binomial outcomes in the fixed sample design and in group sequential designs with O'Brien-Fleming-type and Pocock-type error spending functions.
The black line depicts the nominal one-sided type I error rate of $\alpha=0.025$. The horizontal grey lines mark $\alpha \pm 2SE$ with $SE$ the simulation error at a simulated type I error rate of $0.025$.}
\label{fig:Type1eHF}
\end{figure}

In the following we study the performance of the proposed group sequential procedure for scenarios motivated by the number of heart failure hospitalizations in clinical trials for heart failure patients with preserved ejection fraction.
The scenarios are characterized by large sample sizes of around 1000 patients per arm, small annualized event rates of around 0.1, and unequal exposure times of several years.
The first part of the Monte Carlo simulation study focuses on the assessment of the type I error rate.
We consider the scenarios listed in Table \ref{table:scenarios} which cover a range of shape parameters, rates, sample sizes, and number of data looks in the group sequential design.
In Figure \ref{fig:Type1eHF} the simulated type I error rates of the Wald test for the fixed sample design and the two group sequential designs are displayed.
Each dot represents the simulated type I error rate of a single scenario based on $50\,000$ Monte Carlo replications.
In each frame of the $1\times3$ grid the type I error rate is plotted against the sample size $n_1=n_2$ for a specific design.
For the group sequential designs $n_1=n_2$ refers to the maximum sample size if the trial is not stopped early.
For the fixed sample design fewer scenarios exist because the number of data looks is not varied, i.e., the data is only analyzed at the end of the trial.
Figure \ref{fig:Type1eHF} shows that the Wald test for the group sequential design controls the type I error rate and that no practically relevant type I error inflation compared to the Wald test in the fixed sample design is detectable.
To further assure that the simulation error does not mask a type I error inflation, we performed additional simulations for the scenario with rates $\mu_1 = \mu_2 = 0.14$, a shape parameter of $\phi = 5$, a maximum of $K=5$ data looks, and sample sizes $n_1 = n_2 = 800, \ldots, 1600$.
In this scenario both the overdispersion and the maximum number of data looks are large which is when one would expect a type I error inflation.
The results are reported in Section 1 of the supplemental material and show type I error rates of around $0.0253$ for the fixed sample design and the group sequential design with O'Brien-Fleming-type error spending function.
The type I error rates of the group sequential design with Pocock-type spending function are about $0.02507$.
Due to the large number of $400\,000$ Monte Carlo replications, this assures that no practically relevant type I error inflations are present.\\
Regarding the power, the required maximum information and sample size for a power $1-\beta$ are calculated using the planning approach proposed in the previous section under the assumption of a four year study and an uniform recruitment during an accrual period of 1.25 years.
The simulation results are listed in Table \ref{table:SimResultsHF}.
\begin{sidewaystable}[htb]
\begin{center}
\caption{Results of power related simulations for scenarios motivated by heart failure hospitalizations in chronic heart failure.}
\label{table:SimResultsHF}
\scriptsize
\begin{tabular}{l*{1} {c}{c}{c}{c}{c}l*{1} {c}{c}{c}{c}{c}{c}{c}{c}{c}{c}}
\toprule
& & & & Fix & & & OF & & & & & Pocock & & & & \\
$1-\beta$ & $\theta^{*}$ & $\phi$ & $K$ & $\mathcal{I}_{fix}$ & $n_1$ & Power & $\mathcal{I}_{max}$ & $n_1$ & Power & P(stop early) & $E[\mathcal{I}]$ & $\mathcal{I}_{max}$ & $n_1$ & Power & P(stop early) & $E[\mathcal{I}]$ \\
0.8 & 0.70 & 2 & 2 & 61.74 & 604 & 0.8040 & 61.94 & 606 & 0.8021 & 0.1901 & 57.28 & 69.30 & 678 & 0.8072 & 0.5944 & 52.59 \\
  0.8 & 0.70 & 2 & 3 & 61.74 & 604 & 0.8011 & 62.45 & 611 & 0.7984 & 0.5190 & 53.66 & 72.26 & 707 & 0.8033 & 0.7453 & 50.70 \\
  0.8 & 0.70 & 2 & 5 & 61.74 & 604 & 0.8037 & 63.27 & 619 & 0.7954 & 0.7186 & 51.10 & 74.82 & 732 & 0.8004 & 0.9077 & 49.35 \\
  0.8 & 0.70 & 5 & 2 & 61.71 & 975 & 0.7978 & 61.90 & 978 & 0.7996 & 0.1915 & 57.25 & 69.25 & 1094 & 0.8014 & 0.5940 & 52.77 \\
  0.8 & 0.70 & 5 & 3 & 61.71 & 975 & 0.8011 & 62.47 & 987 & 0.8000 & 0.5220 & 53.57 & 72.22 & 1141 & 0.8002 & 0.7513 & 50.59 \\
  0.8 & 0.70 & 5 & 5 & 61.71 & 975 & 0.8000 & 63.23 & 999 & 0.8004 & 0.7223 & 51.11 & 74.82 & 1182 & 0.7965 & 0.9170 & 49.35 \\
  0.8 & 0.80 & 2 & 2 & 157.59 & 1475 & 0.8008 & 158.23 & 1481 & 0.8027 & 0.1977 & 145.73 & 176.93 & 1656 & 0.8002 & 0.5963 & 134.70 \\
  0.8 & 0.80 & 2 & 3 & 157.59 & 1475 & 0.7990 & 159.62 & 1494 & 0.7983 & 0.5224 & 136.69 & 184.52 & 1727 & 0.8017 & 0.7555 & 128.96 \\
  0.8 & 0.80 & 2 & 5 & 157.59 & 1475 & 0.8030 & 161.55 & 1512 & 0.8009 & 0.7168 & 130.18 & 191.14 & 1789 & 0.8001 & 0.9108 & 125.01 \\
  0.8 & 0.80 & 5 & 2 & 157.61 & 2423 & 0.8014 & 158.19 & 2432 & 0.7963 & 0.2043 & 145.49 & 176.93 & 2720 & 0.8049 & 0.6019 & 134.04 \\
  0.8 & 0.80 & 5 & 3 & 157.61 & 2423 & 0.8028 & 159.63 & 2454 & 0.7996 & 0.5166 & 136.89 & 184.47 & 2836 & 0.8006 & 0.7577 & 129.00 \\
  0.8 & 0.80 & 5 & 5 & 157.61 & 2423 & 0.7957 & 161.51 & 2483 & 0.7982 & 0.7283 & 130.02 & 191.17 & 2939 & 0.7977 & 0.9204 & 125.75 \\
  0.9 & 0.70 & 2 & 2 & 82.59 & 808 & 0.9009 & 82.89 & 811 & 0.9042 & 0.2691 & 72.82 & 91.79 & 898 & 0.9036 & 0.6676 & 64.07 \\
  0.9 & 0.70 & 2 & 3 & 82.59 & 808 & 0.8994 & 83.61 & 818 & 0.9038 & 0.6231 & 67.25 & 95.36 & 933 & 0.9004 & 0.8162 & 59.50 \\
  0.9 & 0.70 & 2 & 5 & 82.59 & 808 & 0.8994 & 84.53 & 827 & 0.9023 & 0.8887 & 62.84 & 98.53 & 964 & 0.8995 & 0.9893 & 56.54 \\
  0.9 & 0.70 & 5 & 2 & 82.60 & 1305 & 0.9009 & 82.86 & 1309 & 0.8997 & 0.2795 & 72.49 & 91.78 & 1450 & 0.9015 & 0.6751 & 63.83 \\
  0.9 & 0.70 & 5 & 3 & 82.60 & 1305 & 0.9024 & 83.55 & 1320 & 0.9011 & 0.6269 & 67.13 & 95.32 & 1506 & 0.8993 & 0.8096 & 59.48 \\
  0.9 & 0.70 & 5 & 5 & 82.60 & 1305 & 0.8987 & 84.50 & 1335 & 0.8999 & 0.8820 & 62.97 & 98.49 & 1556 & 0.8996 & 0.9907 & 56.35 \\
  0.9 & 0.80 & 2 & 2 & 211.01 & 1975 & 0.9033 & 211.76 & 1982 & 0.9008 & 0.2778 & 185.36 & 234.41 & 2194 & 0.9016 & 0.6715 & 163.41 \\
  0.9 & 0.80 & 2 & 3 & 211.01 & 1975 & 0.8995 & 213.47 & 1998 & 0.9024 & 0.6232 & 171.28 & 243.60 & 2280 & 0.8988 & 0.8139 & 152.22 \\
  0.9 & 0.80 & 2 & 5 & 211.01 & 1975 & 0.8976 & 215.93 & 2021 & 0.8973 & 0.8941 & 160.76 & 251.61 & 2355 & 0.8989 & 0.9781 & 144.83 \\
  0.9 & 0.80 & 5 & 2 & 211.01 & 3244 & 0.8992 & 211.73 & 3255 & 0.8992 & 0.2772 & 185.42 & 234.43 & 3604 & 0.9021 & 0.6722 & 163.38 \\
  0.9 & 0.80 & 5 & 3 & 211.01 & 3244 & 0.8989 & 213.55 & 3283 & 0.9035 & 0.6188 & 171.73 & 243.54 & 3744 & 0.8994 & 0.8124 & 152.59 \\
  0.9 & 0.80 & 5 & 5 & 211.01 & 3244 & 0.9021 & 215.89 & 3319 & 0.9013 & 0.8893 & 160.23 & 251.60 & 3868 & 0.9016 & 0.9839 & 143.82 \\
\bottomrule
\end{tabular}
\end{center}
\end{sidewaystable}
Table \ref{table:SimResultsHF} shows that the sample size calculations described in Section \ref{sec:planning} are accurate in the sense that they result in the test having the target power.
In particular, recalculating the critical values at the time of the analysis does not affect the power as long as the information time of the analysis corresponds to the information time considered during the planning phase.
The group sequential designs reduce the expected information compared to a fixed sample design.
Comparing the two group sequential designs, the design with a Pocock-type error spending function results in the larger reduction of the expected information and in the larger probability for rejecting early than the design with O'Brien-Fleming-type error spending function.
However, the group sequential designs required a larger maximum information and a larger maximum sample size compared to the fixed design.
\subsection{MRI lesions in relapsing multiple sclerosis}
\begin{figure}[ht]
\centering
\includegraphics[width=1\textwidth]{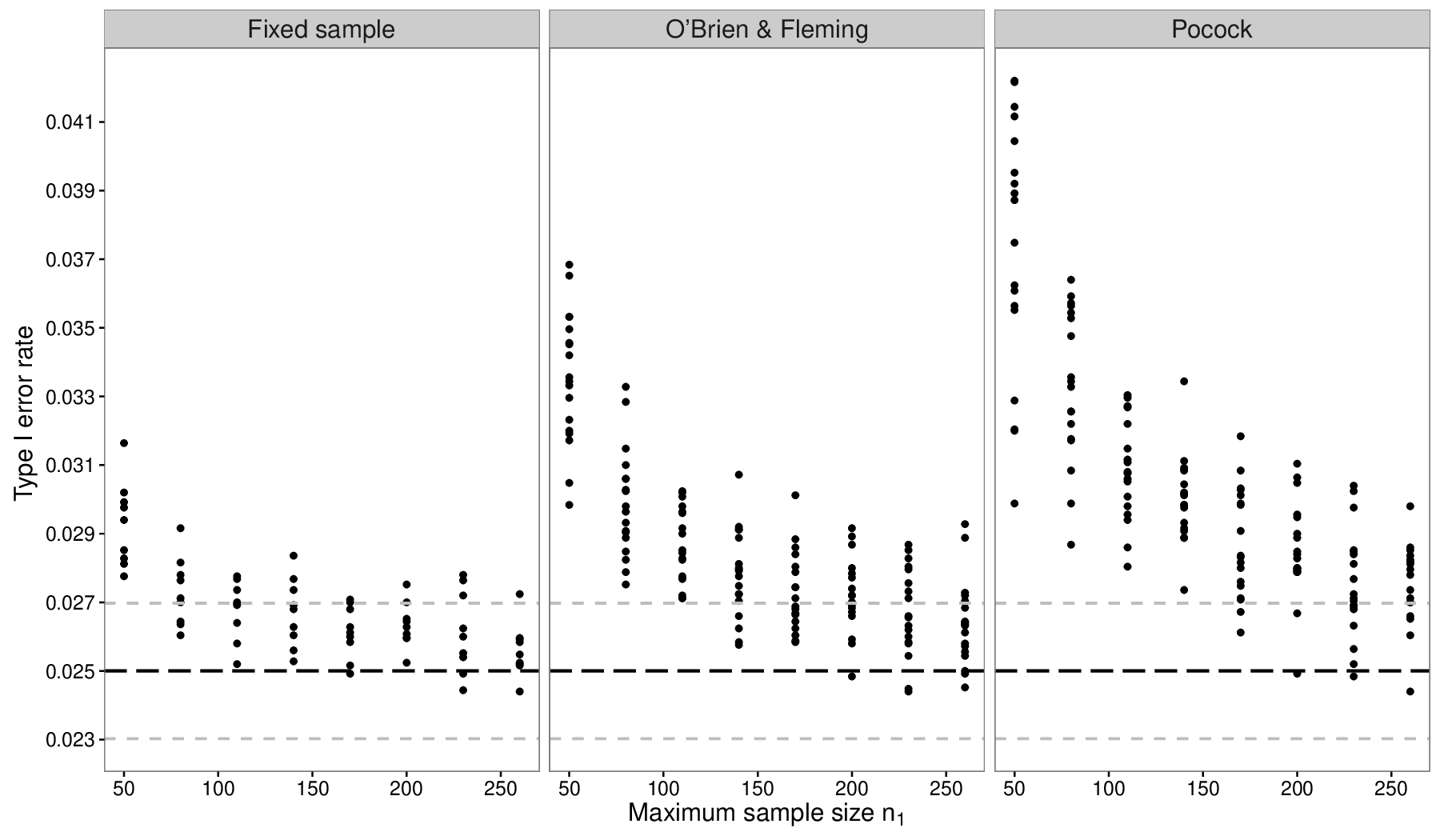}
\caption{Type I error rate of the Wald test for negative binomial outcomes in the fixed sample design and in group sequential designs with O'Brien-Fleming-type and Pocock-type error spending functions.
The black line depicts the nominal type I error rate of $\alpha=0.025$. The horizontal grey lines mark $\alpha \pm 2SE$ with $SE$ the simulation error at a simulated type I error rate of 0.025. }
\label{fig:Type1eMS}
\end{figure}

The simulation scenarios in the previous section were characterized by trial sizes in the thousands, small event rates of less than one event per patient and year, and unequal follow-up times of multiple years.
In contrast, the simulation scenarios in this section are characterized by event rates of around one event per month, equal exposure times of 6 months per patient, and clinical trial sizes of around 100.
The scenarios are motivated by MRI lesion counts in clinical trials in multiple sclerosis.
As in the last paragraph, the results concerning the type I error rate of the Wald test for the fixed sample design and the group sequential design are presented first, followed by the discussion of the power related simulation results.
Figure \ref{fig:Type1eMS} shows the simulated type I error rates. For each rate 25\,000 Monte Carlo replications were performed.
\\
Figure \ref{fig:Type1eMS} shows that even in the fixed sample design the Wald test does not control the type I error rate, which is in agreement with previously published results. \cite{aban2009inferences}
Moreover, the type I error rate of the Wald test in the group sequential designs is inflated compared to the Wald test in the fixed sample design.
The type I error inflation is more extreme in the group sequential design with the Pocock-type error spending function.
The type I error rate of the Wald test in group sequential designs is inflated because in group sequential designs the null hypothesis is tested several times during data look with small sample sizes and the smaller the sample size, the larger the type I error inflation of the Wald test.
However, for every design the type I error rate of the respective Wald test converges to the target type I error rate as the maximum sample size increases. \\ \indent
As in Section \ref{sec:simHF}, the maximum information levels and the sample sizes for the power related simulation study are calculated such that the respective design is powered to the prespecified level $1-\beta$.
The results of the simulation study are listed in Table \ref{table:SimResultsMS}.
\begin{sidewaystable}[htb]
\begin{center}
\caption{Results of power related simulations for scenarios motivated by lesion counts in multiple sclerosis.}
\label{table:SimResultsMS}
\scriptsize
\begin{tabular}{l*{1} {c}{c}{c}{c}{c}l*{1} {c}{c}{c}{c}{c}{c}{c}{c}{c}{c}}
\toprule
& & & & Fix & & & OF & & & & & Pocock & & & & \\
$1-\beta$ & $\theta^{*}$ & $\phi$ & $K$ & $\mathcal{I}_{fix}$ & $n_1$ & Power & $\mathcal{I}_{max}$ & $n_1$ & Power & P(stop early) & $E[\mathcal{I}]$ & $\mathcal{I}_{max}$ & $n_1$ & Power & P(stop early) & $E[\mathcal{I}]$ \\
0.8 & 0.50 & 2 & 2 & 16.33 & 77 & 0.8002 & 16.33 & 77 & 0.8043 & 0.2589 & 15.04 & 18.24 & 86 & 0.8057 & 0.6032 & 14.18 \\
  0.8 & 0.50 & 2 & 3 & 16.33 & 77 & 0.8050 & 16.55 & 78 & 0.8034 & 0.5420 & 14.27 & 19.09 & 90 & 0.8052 & 0.7638 & 13.49 \\
  0.8 & 0.50 & 3 & 2 & 16.38 & 110 & 0.7989 & 16.38 & 110 & 0.8047 & 0.2413 & 15.08 & 18.32 & 123 & 0.8067 & 0.6069 & 14.13 \\
  0.8 & 0.50 & 3 & 3 & 16.38 & 110 & 0.7983 & 16.53 & 111 & 0.8017 & 0.5455 & 14.16 & 19.06 & 128 & 0.8049 & 0.7577 & 13.44 \\
  0.8 & 0.70 & 2 & 2 & 61.60 & 282 & 0.8004 & 62.03 & 284 & 0.8009 & 0.2155 & 57.09 & 69.24 & 317 & 0.8029 & 0.5949 & 53.08 \\
  0.8 & 0.70 & 2 & 3 & 61.60 & 282 & 0.8019 & 62.47 & 286 & 0.8017 & 0.5309 & 53.43 & 72.30 & 331 & 0.8018 & 0.7556 & 50.73 \\
  0.8 & 0.70 & 3 & 2 & 61.72 & 406 & 0.8012 & 61.87 & 407 & 0.8002 & 0.2149 & 56.85 & 69.32 & 456 & 0.8057 & 0.5998 & 52.85 \\
  0.8 & 0.70 & 3 & 3 & 61.72 & 406 & 0.7946 & 62.48 & 411 & 0.8034 & 0.5258 & 53.42 & 72.21 & 475 & 0.7970 & 0.7553 & 50.79 \\
  0.9 & 0.50 & 2 & 2 & 21.85 & 103 & 0.9010 & 21.85 & 103 & 0.9003 & 0.3142 & 19.15 & 24.39 & 115 & 0.9046 & 0.6816 & 17.23 \\
  0.9 & 0.50 & 2 & 3 & 21.85 & 103 & 0.9012 & 22.06 & 104 & 0.8994 & 0.6260 & 17.81 & 25.24 & 119 & 0.9030 & 0.8234 & 15.90 \\
  0.9 & 0.50 & 3 & 2 & 21.89 & 147 & 0.8999 & 21.89 & 147 & 0.9052 & 0.3066 & 19.17 & 24.28 & 163 & 0.9031 & 0.6739 & 17.16 \\
  0.9 & 0.50 & 3 & 3 & 21.89 & 147 & 0.9028 & 22.19 & 149 & 0.9035 & 0.6293 & 17.83 & 25.17 & 169 & 0.9008 & 0.8194 & 15.83 \\
  0.9 & 0.70 & 2 & 2 & 82.56 & 378 & 0.8996 & 82.78 & 379 & 0.8996 & 0.2890 & 72.42 & 91.74 & 420 & 0.9001 & 0.6699 & 64.44 \\
  0.9 & 0.70 & 2 & 3 & 82.56 & 378 & 0.9017 & 83.66 & 383 & 0.9007 & 0.6241 & 67.18 & 95.23 & 436 & 0.9038 & 0.8177 & 59.42 \\
  0.9 & 0.70 & 3 & 2 & 82.54 & 543 & 0.9004 & 82.85 & 545 & 0.9014 & 0.2895 & 72.30 & 91.82 & 604 & 0.9019 & 0.6729 & 64.25 \\
  0.9 & 0.70 & 3 & 3 & 82.54 & 543 & 0.8982 & 83.61 & 550 & 0.9039 & 0.6244 & 67.02 & 95.31 & 627 & 0.9012 & 0.8130 & 59.53 \\
\bottomrule
\end{tabular}
\end{center}
\end{sidewaystable}
Table \ref{table:SimResultsMS} shows that the approach for planning group sequential designs presented in Section \ref{sec:planning} already works for maximum  sample sizes of around 80 patients per treatment arm although the planning method relies on asymptotic considerations.
However, it is important to keep in mind that the type I error rate is not controlled for smaller sample sizes.
The relationship between the maximum information for the fixed sample design and for the group sequential design is similar to the scenarios presented in Table \ref{table:SimResultsHF}.
\section{Adjustments for small sample sizes}
\label{sec:smallSample}
The type I error rate of the proposed Wald test for group sequential designs is inflated in the case of small sample sizes as illustrated in Figure \ref{fig:Type1eMS}.
The type I error rate inflation is the result of a mismatch between the distribution of the Wald statistic for negative binomial outcomes and the assumed multivariate normal distribution when calculating the critical values.
By means of simulations, it can be observed that the distribution of the Wald statistic for small sample sizes has heavier tails than a normal distribution.
Thus, the type I error rate inflation of the Wald test can be reduced by adjusting the Wald statistic such that its distribution is approximated more closely by a normal distribution or by adjusting the calculation of the critical values.
In this section we suggest modifications of the Wald statistic and the critical values which reduce the type I error rate inflation of the Wald test for group sequential designs with negative binomial outcomes.
It is important to emphasize that while we discuss adjustments to the Wald statistics and adjustments to the critical values in separate subsections, both adjustment can be applied simultaneously to the initially proposed Wald test for group sequential designs.
\subsection{Adjusting the Wald statistic}
To obtain a Wald statistic which is approximated more closely by a normal distribution, we propose to estimate the variance in the Wald statistic restricted to the parameter space of the null hypothesis.
This has already been shown to reduce the type I error inflation in several designs, among them also designs with negative binomial outcomes.\cite{farrington1990test,tang2006sample,mutze2016design}
In detail, in the context of group sequential designs the variance estimator for the Wald statistic $T_k$ at look $k$ is an estimator for the information level $\mathcal{I}_{k}$.
The restricted variance estimator, i.e., an information level estimator restricted to the null hypothesis, is obtained by plugging in the restricted maximum likelihood estimators of the rates and the shape parameter into \eqref{eq:infoK}.
With $L(\beta_1,\beta_2,\phi|Y_{ijk}, j = 1,\ldots, n_i, i=1,2)$ the likelihood function at look $k$, the maximum likelihood estimators restricted to the null hypothesis $H_0$ are defined by
\begin{align*}
(\hat{\beta}_{1k}^{(R)}, \hat{\beta}_{2k}^{(R)}, \hat{\phi}_{1k}^{(R)})= \underset{\beta_1 - \beta_2 \geq \log(\delta)}{\operatorname{arg\,max}} \,\, L(\beta_1,\beta_2,\phi|Y_{ijk}, j = 1,\ldots, n_i, i=1,2).
\end{align*}
The restricted maximum likelihood estimators and the restricted information level estimator are consistent estimators under the null hypothesis and, thus, the results about the asymptotic normality of the joint distribution of the Wald statistics under the null hypothesis still hold.
\subsection{Modifying the critical values}
To limit the type I error rate inflation in the case of small sample sizes, we discuss several approaches for modifying the critical values.
\paragraph{Pocock modification}
Pocock (1977) suggested to calculate the critical values for a group sequential design with Student's t-test by transforming the critical values from the normal distribution with the inverse cumulative distribution function of a t-distribution, that is the distribution of the test statistic.\cite{pocock1977group}
In general, the idea is to obtain the critical value for data look $k$ through the transformation $F^{-1}(\Phi(c_k))$, where $c_k$ is the critical value obtained through the canonical joint distribution, $\Phi(\cdot)$ the cumulative distribution function of the standard normal distribution, and $F^{-1}(\cdot)$ the inverse cumulative distribution function of the distribution of the test statistic.\cite{jennison2000group} \\ \indent
The finite sample size distribution of the Wald statistic $T_k$ is unknown.
As an approximation, $F(\cdot)$ can be chosen as the cumulative distribution of Student's t-distribution, because Student's t-distribution has heavier tails than a normal distribution and approximates the distribution of the Wald statistics better than the normal distribution for small sample sizes. \cite{mutze2017studentized}
Alternatively, $F(\cdot)$ can be the cumulative distribution function of a resampling distribution.
In the following we outline the use of the permutation distribution of the Wald statistics for calculating the critical values.
For this purpose, let $(\tau(i))_{i\leq n}$ be a random variable which is uniformly distributed on the group $S_n$, i.e., the group of all permutations of the vector $(1,2,\ldots,n)$, and let $\tilde{\mathbb{P}}$ be the respective probability measure.
Let $\mathbb{Y}_k=(Y_{11k}, \ldots, Y_{2{n_2}k})$  be the vector of the random variables at data look $k$ and let $\tau(\mathbb{Y}_k)$ denote the randomly permuted vector $\mathbb{Y}_k$.
The probability measure $\tilde{\mathbb{P}}$ is independent of the vector $\mathbb{Y}_k$.
With $T(\mathbb{Y}_k)$ the Wald statistic as a function of the random variables, the permutation statistic is defined by results of the mapping
\begin{align*}
(\tau(i))_{i\leq n} \mapsto T(\tau(\mathbb{Y}_k)) | \mathbb{Y}_k  .
\end{align*}
Then, the permutation distribution refers to the distribution of the permutation statistic $T(\tau(\mathbb{Y}_k)) | \mathbb{Y}_k$ with respect to the probability measure $\tilde{\mathbb{P}}$.
For the Pocock modification of the critical values in group sequential testing, we propose to choose  $F(\cdot)$ as the cumulative distribution function of the permutation distribution.
\paragraph{Multivariate t-distribution}
Since the Wald statistic for negative binomial outcomes has heavier tails than the normal distribution but is still symmetric around zero under the null hypothesis, it seems natural to approximate the joint distribution of the Wald statistics $T_1,\ldots, T_K$ by a multivariate Student's t-distribution.
This means that the critical value $c_k$ for the Wald test at look $k$ is calculated by solving \eqref{eq:criticCal} under the assumption that the joint distribution of test statistics $T_1,\ldots, T_k$ follows a multivariate Student's t-distribution.
The multivariate Student's t-distribution has a mean vector $0$, the same $k\times k$-dimensional covariance matrix $\Sigma$ as defined in \eqref{eq:jointdistCov}, and $\nu$ degrees of freedom.
There is a variety of potential choices for the number of degrees of freedom $\nu$ of the multivariate  Student's t-distribution.
In the context of testing treatment differences in nested subgroups, which are statistically the same problem as testing in group sequential designs as illustrated by Spiessens and Debois (2010). \cite{spiessens2010adjusted}
Placzek and Friede (2017) \cite{placzek2017clinical} proposed a conservative and a liberal choice for the degrees of freedom of the multivariate Student's t-distribution.
The conservative choice is to use the degrees of freedom of the smallest sub population and the liberal choice is to use the degrees of freedom of the full population.
Transferred to group sequential designs, the conservative choice for the degrees of freedom are the degrees of freedom of the first data look and the liberal choice are the (planned) degrees of freedom of the last data look.
As noted by Graf \textit{et al.} (2017) the Pocock approach with $F(\cdot)$ the cumulative distribution function of Student's t-distribution and calculating the critical values through a multivariate Student's t-distribution do not result in the same critical values.\cite{graf2017robustness}  \\ \indent
\subsection{Simulation study}
\begin{figure}[ht]
\centering
\includegraphics[width=1\textwidth]{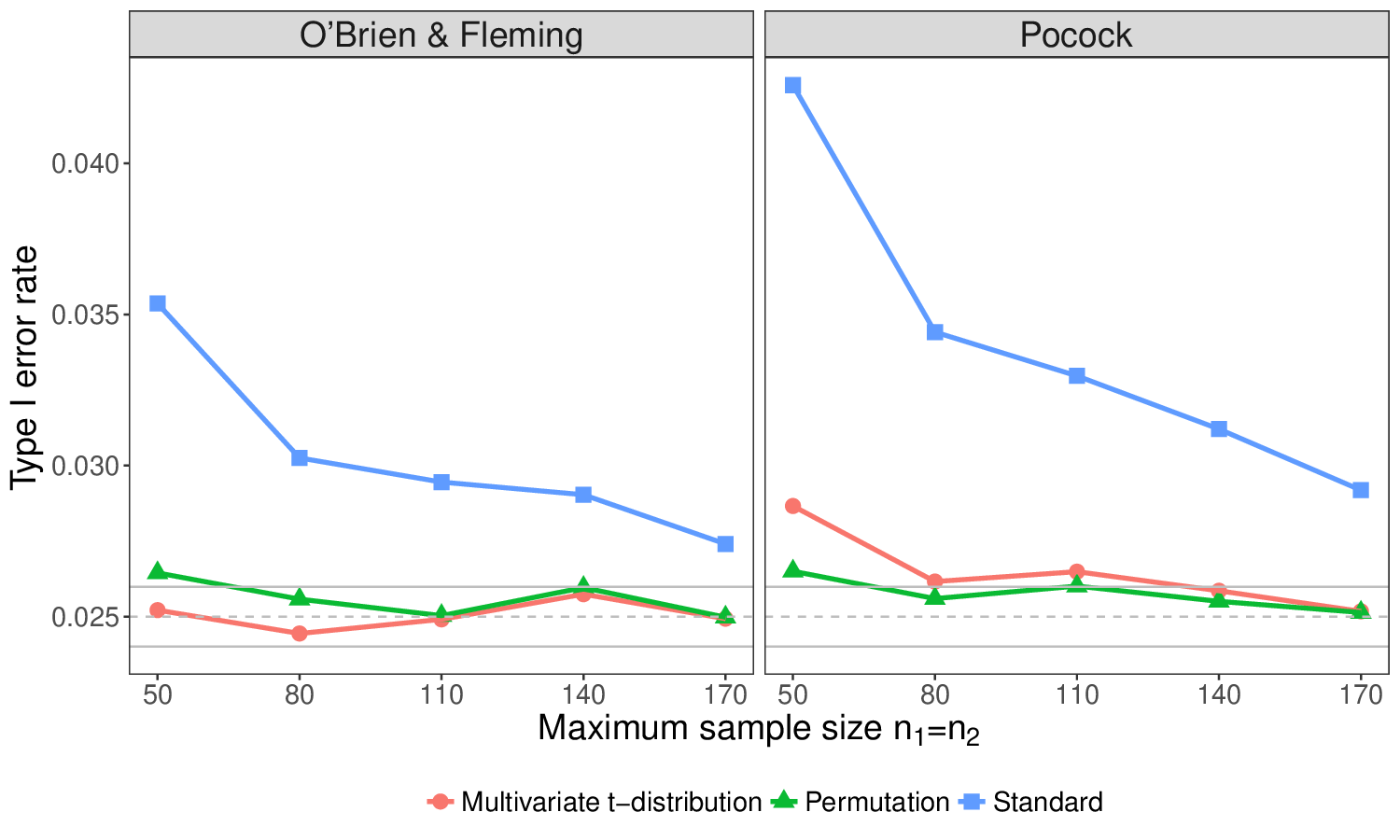}
\caption{Type I error rate of the Wald test with and without modifications for negative binomial outcomes in group sequential designs with O'Brien-Fleming-type and Pocock-type error spending function.
The black line depicts the planned one-sided type I error rate of $\alpha=0.025$. The horizontal grey lines mark $\alpha \pm 2SE$ with $SE$ the simulation error at a simulated type I error rate of 0.025.}
\label{fig:smallsample}
\end{figure}

In this subsection we assess the performance of the proposed adjustments to the Wald statistics and the critical values in group sequential designs with negative binomial outcomes.
Since both adjustments can be applied simultaneously or separately, there are several group sequential testing procedures worth investigating.
We restrict our presentation of the results to the two procedures which actually resulted in an appropriate control of the type I error rate.
The first procedure tests based on the Wald statistic with the variance estimation restricted to the null hypothesis and the critical values are calculated using the multivariate t-distribution with the degrees of freedom the number of subjects recruited at the first data look.
The second procedure also tests the null hypothesis using the Wald statistic with the variance estimation restricted to the null hypothesis and its critical values are based on Pocock's modification with the permutation distribution as the cumulative distribution function $F(\cdot)$.
The number of permutations of a vector with length $n$ is too large to exactly determine the permutation distribution.
Instead, for a given vector $\mathbb{Y}_k$, $15\,000$ random permutations of the vector are generated and for each of the resulting permuted vector the test statistic $T(\tau(\mathbb{Y}_k))$ is calculated.
For calculating the critical value, the resulting empirical cumulative distribution function $\hat{F}(\cdot)$ is utilized.
Since the second procedure requires intensive computations, we only vary the sample size and keep the remaining parameters fix, i.e., we choose annualized rates $\mu_1 = \mu_2 = 10$, shape parameter $\phi = 4$, number of data looks $K = 3$, and the other parameters as listed in Table \ref{table:scenarios}.
This scenario is characterized by a large overdispersion which is associated with a large type I error inflation of our initially proposed group sequential procedure.
Therefore, the procedure which controls the type I error rate for the considered scenario also control the type I error rate for scenarios in which the initially proposed procedure has a smaller type I error inflation.
The results are illustrated in Figure \ref{fig:smallsample}.
Figure \ref{fig:smallsample} shows that both modifications of the group sequential procedure yield type I error rates close to the target level $\alpha = 0.025$.
For the O'Brien-Fleming-type error spending function, the modification based on the multivariate Student's t-distribution results in type I error rates closer to the nominal level than the procedure based on Pocock's modification with the permutation distribution as the cumulative distribution function.
For the Pocock-type spending function, these results are reversed, i.e., the procedure based on Pocock's modification performs better.\\
Noteworthy, the group sequential procedure based on the Pocock's modification with the permutation distribution is not to be confused with the actual group sequential permutation test.
The group sequential permutation test calculates the critical values as outlined in Section \ref{sec:groupseq} based on a joint distribution obtained through resampling instead of based on a multivariate normal distribution.
In more detail, at data look $\tilde{k}$ the joint permutation distribution is obtained by permuting the treatment indicators and then calculating the test statistics $(T_1,\ldots, T_{\tilde{k}})$ from different data looks. In practice, the number of permutations is too large to actually calculate the test statistics $(T_1,\ldots, T_{\tilde{k}})$ for all available permutations.
Thus, the permutation distribution is approximated by calculating the test statistics for a set of random permutations.
A group sequential permutation test will control the type I error rate but is also computationally more expensive than Pocock's modification with the permutation distribution.
While this is not a problem when analyzing a single trial, this can be an issue when evaluating permutation tests in simulation studies, even if efficient methods for assessing permutation tests, see  Jennison (1992),\cite{jennison1992} are used.
\section{Clinical trial examples revisited}
\label{sec:examplesrev}
In this section we study the effects of rejecting early in a group sequential design.
We illustrate how an early rejection effects the sample size, study duration, information level, and total follow-up years.
The parameters used in our illustration are motivated by the clinical trial examples from Section \ref{sec:Examples}.
The conclusions drawn from the examples cannot be generalized and are specific to the design parameters.
\subsection{Chronic heart failure}
\begin{figure}[ht]
\centering
\includegraphics[width=1\textwidth]{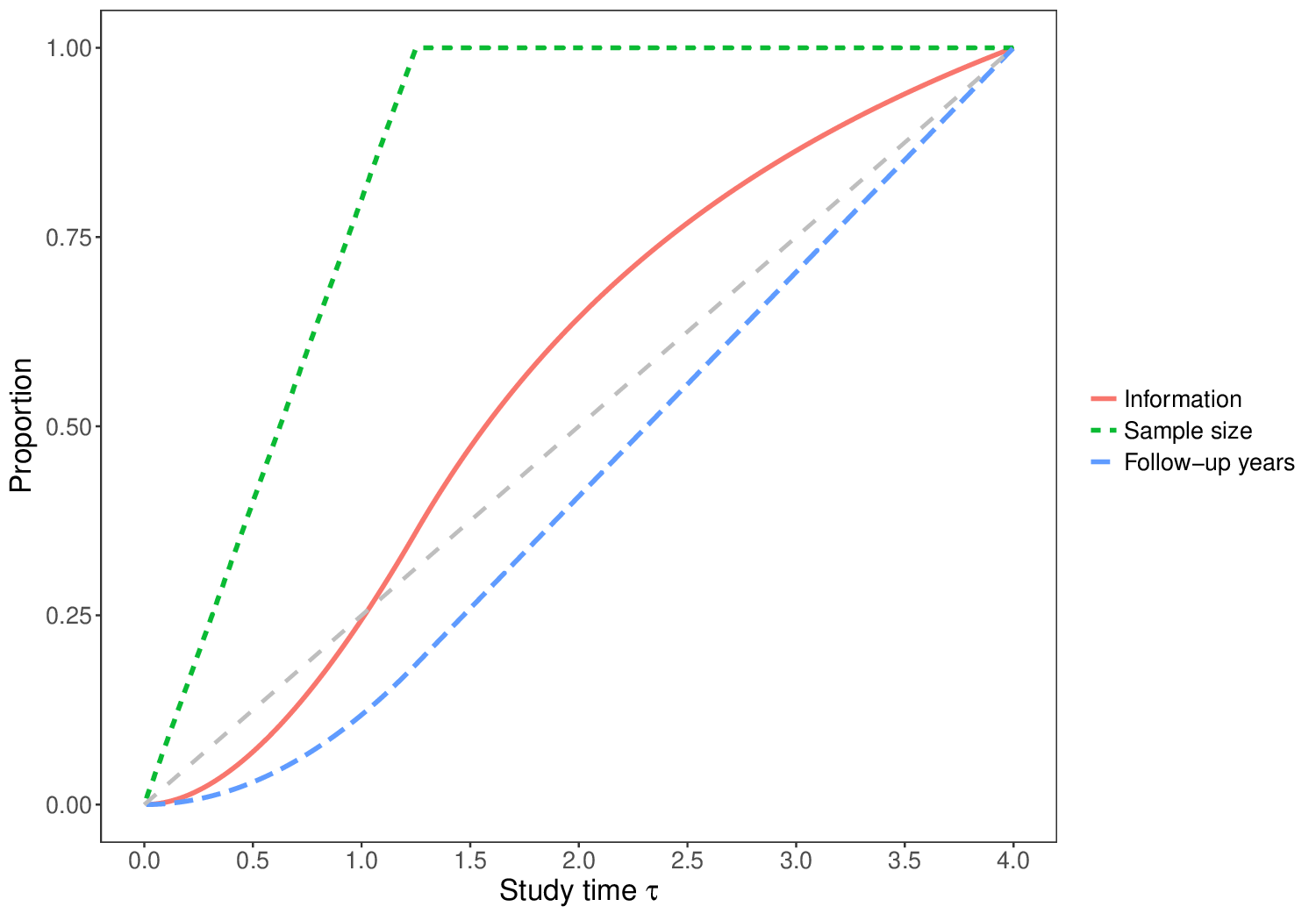}
\caption{Proportions $\mathcal{I}_{\tau}/\mathcal{I}_{max}$, $n_{\tau}/n_{max}$, and  $t_{\tau}/t_{max}$ versus the study time $\tau$. The dashed grey line marks a linear increase in the study time $\tau$. For the scenario considered, see text.}
\label{fig:recruitHF}
\end{figure}

We consider a clinical trial which is planned with a group sequential design with O'Brien-Fleming-type error spending function and one interim analysis at $0.5\mathcal{I}_{max}$, powered to $80\%$ for the alternative $\theta^{*}=0.7$ under the assumption of the nuisance parameters $\phi=5$ and $\mu_{2}=0.125$.
Moreover, the clinical trial is planned for a duration of four years, with an accrual period of 15 months, and a planned recruitment of 1956 patients (in total).
The recruitment is assumed to be uniform in the accrual period.
Let $\tau$ be the time since recruiting the first patient and let $\mathcal{I}_{\tau}$, $n_{\tau}$, and $t_{\tau}$ be the information level, the total sample size, and the total follow-up years at time $\tau$.
Additionally, let $\mathcal{I}_{max}$, $n_{max}$, and $t_{max}$ be the respective parameter at the end of study if the study is not stopped early for efficacy.
In Figure \ref{fig:recruitHF} the proportions $\mathcal{I}_{\tau}/\mathcal{I}_{max}$, $n_{\tau}/n_{max}$, and  $t_{\tau}/t_{max}$
are plotted against the time $\tau$.
Figure \ref{fig:recruitHF} highlights that the information level, the sample size, and the total follow-up years are not linear in the study duration $\tau$.
Moreover, at the end of the accrual period, that is after $\tau = 1.25$ years, only $36\%$ of the maximum information level $\mathcal{I}_{max}$ has been accumulated.
Thus, in this clinical trial with a group sequential design where the interim analysis is performed at an information time of $0.5\mathcal{I}_{max}$, the number of patients would not be lowered if  the trial is stopped early.
The benefit of an early stopping is shortening the clinical trial by over two years.
Moreover, Figure \ref{fig:recruitHF} depicts that the longer the trial continues, the smaller the information gain per time unit about the unknown treatment effect.
\subsection{Relapsing-remitting multiple sclerosis}
\begin{figure}[ht]
\centering
\includegraphics[width=1\textwidth]{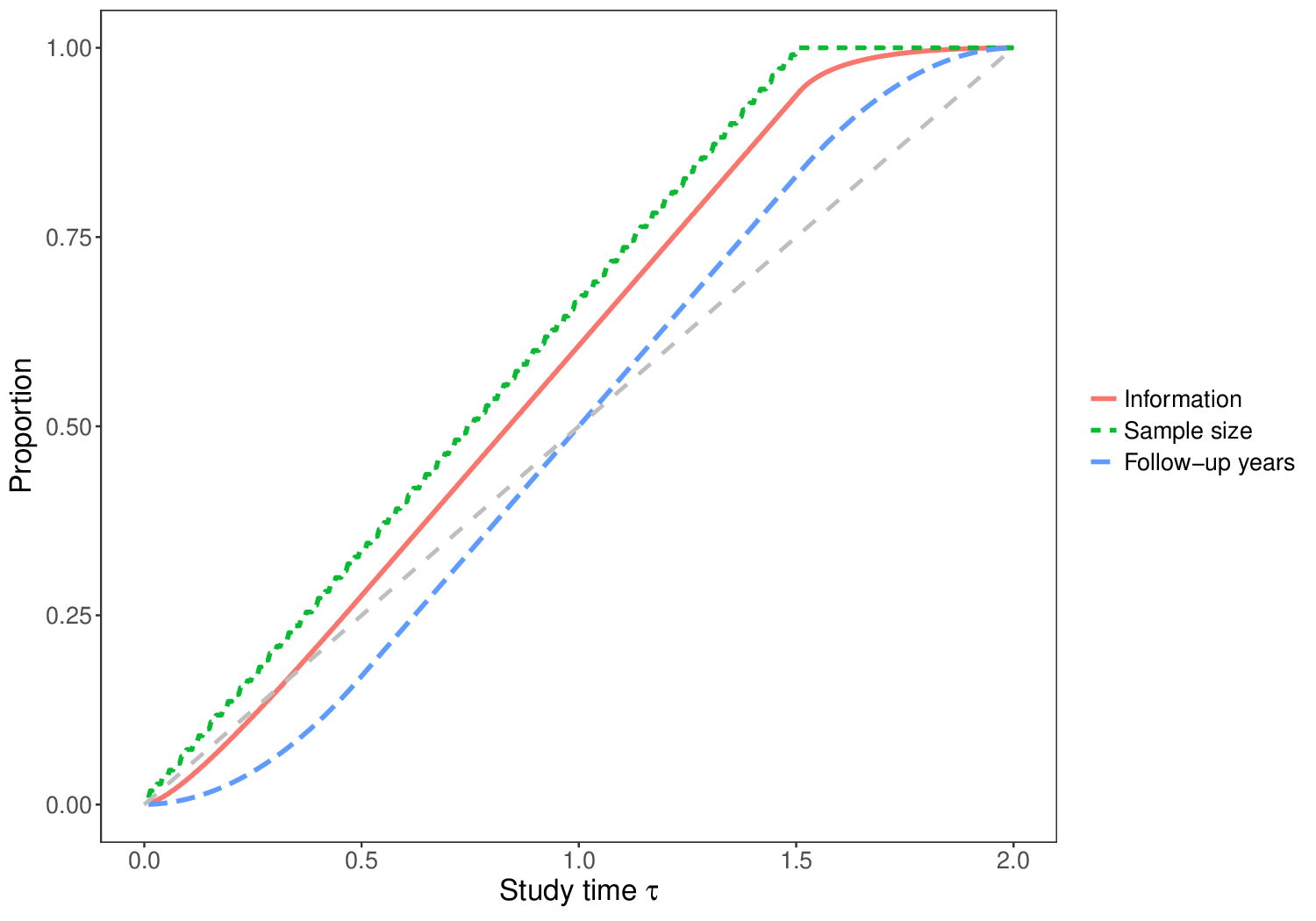}
\caption{Proportions $\mathcal{I}_{\tau}/\mathcal{I}_{max}$, $n_{\tau}/n_{max}$, and  $t_{\tau}/t_{max}$ versus the study time $\tau$ in years. The dashed grey line marks a linear increase of the proportions in the study time $\tau$. For the scenario considered, see text.}
\label{fig:recruitMS}
\end{figure}

Motivated by the clinical trial in multiple sclerosis discussed in Section \ref{sec:Examples}, in the following, the focus is on a clinical trial where the accrual period is 18 months and each patient has an identical follow-up time of 6 months.
The patients are recruited uniformly throughout the accrual period.
The clinical trial has a group sequential design with O'Brien-Fleming-type error spending function with one interim analysis.
In total, 220 patients are included into the trial which corresponds to a power of $80\%$ for a rate ratio $\theta^{*}=0.5$ and nuisance parameters $\phi=3$ and $\mu_2=8.4$.
The sample size of a fixed sample design with the same parameters is identical.
In Figure \ref{fig:recruitMS} the proportions $\mathcal{I}_{\tau}/\mathcal{I}_{max}$, $n_{\tau}/n_{max}$, and  $t_{\tau}/t_{max}$ are plotted against the study time $\tau$.
Figure \ref{fig:recruitMS} highlights that in the studied setting of identical follow-up times of six months much can be gained with respect to sample size reduction by conducting a group sequential design instead of a fixed sample design.
For example, if the trial is stopped after an interim analysis conducted at an information time of $\mathcal{I}_{\tau}/\mathcal{I}_{max}=0.5$, which is achieved at $\tau = 0.84$ years, the sample size would be reduced from 220 to 122 patients.
Compared to the example of a clinical trial in heart failure with multiple years of individual follow-up times, the potential for reducing the trial duration is limited since the trial runs for two years and recruiting is planned for 1.5 years.
\section{Discussion and outlook}
\label{sec:discussion}
In this manuscript we proposed a Wald test based on maximum likelihood estimates for analyzing group sequential designs with negative binomial outcomes.
The proposed testing procedure asymptotically follows the canonical joint distribution, that is the increments are asymptotically independent.
In a simulation study with scenarios motivated by clinical trials in chronic heart failure and relapsing multiple sclerosis we assessed the finite sample size characteristics of the proposed testing procedure.
The simulation study shows that the testing procedure controls the type I error rate for large sample sizes but also revealed that the testing procedure does not control the type I error rate for small sample sizes and high overdispersion.
The magnitude of the type I error inflation in the group sequential design compared to the fixed sample design depends on the number of data looks, the sample size, the overdispersion, and on the error spending function.
In Section \ref{sec:smallSample} we suggested modifications of the proposed Wald test which resulted in type I error rates closer to the nominal level.
From the results of the simulation study we can also conclude that the planning methodology already performs well for small maximum sample sizes of around 80 patients per arm even for large overdispersion indices between 6 and 20.
The proposed methodology for planning group sequential designs with negative binomial outcomes is based on the theory of the canonical joint distribution.
However, the information level for negative binomial outcomes depends on the distribution parameters, i.e.,  the rates and the shape parameter, the sample size, but also on the exposure times of the individual patients.
We discussed in detail the resulting flexibility for group sequential designs with negative binomial outcomes in Section \ref{sec:examplesrev}.\\ \indent
Group sequential designs with negative binomial outcomes are so far not implemented in any software package for group sequential designs.
Our new R package \textit{gscounts} implements the methodology for planning and analyzing group sequential design with negative binomial outcomes as described in this manuscript.\cite{github2017gscounts}
The implementation includes the option for binding and non-binding futility stopping.
Additionally, the R package \textit{gscounts} also contains functions for planning fixed sample designs with negative binomial outcomes.\\ \indent
In this article, we considered group sequential designs, where stopping for success is based on a statistical significance test.
In early phase trials, often double criteria for stopping for success are used, which are based on both statistical significance and clinical relevance.\cite{kieser2005assessment,fisch2015bayesian}
Group sequential designs using such double criteria are described by Gsponer \textit{et al.}, which also discuss a multiple sclerosis example with negative binomially distributed lesion count data.\cite{gsponer2014practical}\\ \indent
The negative binomial model is a common choice to model MRI lesions in relapsing-remitting multiple sclerosis as discussed in Section \ref{sec:Examples}.
However, since the number of MRI lesions is gathered based on MRIs which are performed every couple of weeks or months during a clinical trial, the true time points at which new or enlarging MRI lesions occur is unknown.
Instead, generally only the time point at which the MRI was performed and how many new or enlarging lesions were counted are known.
Asendorf \textit{et al.} proposed a negative binomial integrated value autoregressive process for modeling MRI lesions in this situation. \cite{asendorf2017}
As an extension of the methods presented in this manuscript, group sequential designs could be studied for negative binomial integrated value autoregressive processes.\\ \indent
A covariate is typically included into the analysis of clinical trials if the covariate has a strong impact on the primary outcome.
The aim when including a covariate is to minimize the bias and maximize the efficiency of the statistical analysis. \cite{CHMPcovariates}
Covariate adjustment in group sequential clinical trials has already been studied for normal linear models. \cite{jennison1997group}
The natural extension of the negative binomial distribution considered in this manuscript to a model with covariates is the negative binomial regression model. \cite{lawless1987negative}\\ \indent
When the count data cannot be modeled with a parameteric distribution often quasi-Poisson models are applied.
In quasi-Poisson models, the rates are estimated with a method of moments estimator. \cite{mccullagh1989generalized}
Test statistics based on these method of moments estimators do in general not fulfill the independent increment assumption.
Thus, additional methodological work is required to extend the already existing nonparametric and semiparametric methods for group sequential designs with recurrent events by methods for the the quasi-Poisson model. \\ \indent
In the simulation study presented in this manuscript, the data looks were performed at prespecified calendar times which in turn were determined based on the true parameters.
In practice the information on the nuisance parameters and the accrual rate are often not reliable enough to accurately determine the calendar time at which a certain information level is attained prior to the trial.
Therefore, clinical trials are monitored and the time point of a data look is often selected based on the available data.
Monitoring procedures and the potential effect on the type I error rate have already been studied extensively. \cite{proschan1992effects, scharfstein1998use, mehta2001flexible}
As an alternative to classical frequentist procedures a Bayesian approach has been suggested. \cite{parmar2001monitoring}
In future research procedures for monitoring group sequential designs with negative binomial outcomes should be developed and their operating characteristics be assessed.
In particular, if the information of a trial is monitored and the trial is stopped at a prespecified maximum information, design aspects such as the sample size and study duration can or must be adaptable.
Thus, part of the development of monitoring procedures must be the assessment of design adaptation rules; for the sample size existing approaches can be applied. \cite{friede2010blindeda,friede2010blindedb} \\ \indent
Testing nested subgroups is statistically the same problem as testing in a group sequential design  as illustrated by Spiessens and Debois. \cite{spiessens2010adjusted}
Therefore, the results presented in this manuscript can also be applied to clinical trials with nested subgroups and are relevant to precision and personalized medicine.
\appendix
\section{Maximum likelihood estimators}
\label{sec:ApendA}
In this appendix we show that at data look $k=2,\ldots, K$ the negative binomially distributed random variables $Y_{ijk}$ $(j = 1, \ldots, n_i, i = 1,2)$ are sufficient for $(\beta_1, \beta_2, \phi)$, i.e.,  that the distribution of $Y_{ij1}, \ldots, Y_{ij(k-1)} | Y_{ijk}$ does not depend on the rates or the shape parameter.
Therefore, the maximum likelihood estimators for the rates and the shape parameter at data look $k$ only depends on the random variables $Y_{ijk}$  $(j = 1, \ldots, n_i, i = 1,2)$ and not on the counts from previous data looks.\\ \indent
For independently Poisson distributed random variables $X_i\sim \operatorname{Pois}(\lambda_i)$ with $i=1,\ldots, n$, conditioned on their sum, the vector of Poisson random variables $X_1,\ldots, X_n$ follows a multinomial distribution:
\begin{align*}
  X_1,\ldots, X_n | X_{\cdot} \sim \operatorname{Multinom}\left(n, \mathbf{p}\right).
\end{align*}
Here, $X_{\cdot} = \sum\limits_{i = 1}^{n}X_i$ is the sum of random variables and the \textit{i}-th entry of vector $\mathbf{p}\in \mathbb{R}^n$ is defined by $p_i = \lambda_i / \lambda_{\cdot}$ with $\lambda_{\cdot}$ the sum of the rates $\lambda_i$ $(i=1,\ldots, n)$.
In the following we use this property about independently Poisson distributed random variables to show that at data look $k=2,\ldots, K$ the random variables $Y_{ijk}$ $(j = 1, \ldots, n_i, i = 1,2)$ are sufficient for $(\beta_1, \beta_2, \phi)$.
Conditioned on the subject specific rate $\lambda_{ij}$, the increments $Y_{ij1},  Y_{ij2}-Y_{ij1},\ldots, Y_{ijk} - Y_{ij(k-1)}$ are independent and Poisson distributed.
With the above property for independent Poisson distributions, it follows that conditioned on the rate $\lambda_{ij}$ and on the random variable $Y_{ijk}$, the vector of increments follows a multinomial distribution, i.e.,
\begin{align*}
  & Y_{ij1},  Y_{ij2}-Y_{ij1},\ldots, Y_{ijk} - Y_{ij(k-1)} \Big\vert Y_{ijk}, \lambda_{ij}
  \sim
  \operatorname{Multinom}\left(n, \mathbf{p}\right),\\
  & \mathbf{p} = \left(\frac{t_{ij1}}{t_{ijk}},\frac{t_{ij2}-t_{ij1}}{t_{ijk}}, \ldots, \frac{t_{ijk}-t_{ij(k-1)}}{t_{ijk}}\right).
\end{align*}
From the conditional joint distribution of the increments, it follows that the distribution of $Y_{ij1}, \ldots, Y_{ij(k-1)} | Y_{ijk}, \lambda_{ij}$ does not depend on $ \lambda_{ij}$ and therefore it also does not depend on the parameters $(\beta_1, \beta_2, \phi)$:
\begin{align*}
  & \mathbb{P}\left(
    Y_{ij1} = y_{ij1}, \ldots, Y_{ij(k-1)} = y_{ij(k-1)} | Y_{ijk} = y_{ijk}, \lambda_{ij}
  \right) \\
  = &
  \mathbb{P}\Big(
    Y_{ij1} = y_{ij1}, Y_{ij2} - Y_{ij1} = y_{ij2} - y_{ij1}, \ldots, \\
    & \qquad Y_{ijk} - Y_{ij(k-1)} = y_{ijk}-y_{ij(k-1)} | Y_{ijk} = y_{ijk}, \lambda_{ij}
  \Big)\\
  = & \frac{y_{ijk}!}{y_{ij1}!\cdot (y_{ij2} - y_{ij1})! \cdot \ldots \cdot (y_{ijk} - y_{ij(k-1)})!} \\
  & \times
  \left(\frac{t_{ij1}}{t_{ijk}}\right)^{y_{ij1}}
  \left(\frac{t_{ij2}-t_{ij1}}{t_{ijk}}\right)^{y_{ij2} - y_{ij1}} \ldots \left(\frac{t_{ijk}-t_{ij(k-1)}}{t_{ijk}}\right)^{y_{ijk} - y_{ij(k-1)}}
  .
\end{align*}
Since the distribution of $Y_{ij1}, \ldots, Y_{ij(k-1)} | Y_{ijk}, \lambda_{ij}$ is independent of the subject specific event rate $\lambda_{ij}$, the marginal distribution $Y_{ij1}, \ldots, Y_{ij(k-1)} | Y_{ijk}$  is also independent of $\lambda_{ij}$.
This makes the random variable $Y_{ijk}$ a sufficient statistic for the parameters $(\beta_1, \beta_2, \phi)$ at data look $k$. \\ \indent
Next, we outline the maximum likelihood estimators for the parameters $(\beta_1, \beta_2, \phi)$. For the sake of readability, we omit the third index $k$ in all your notations.
The maximum likelihood estimators $(\hat{\beta}_{1}, \hat{\beta}_{2}, \hat{\phi})$ for the log-rates and the shape parameter are the solutions to the following equations \cite{lawless1987negative}
\begin{align*}
&\sum_{j=1}^{n_1}\frac{Y_{1j} - t_{1j}\exp(\beta_1)}{1 + \phi t_{1j}\exp(\beta_1)}  = 0,\\
&\sum_{j=1}^{n_2}\frac{Y_{2j} - t_{2j}\exp(\beta_2)}{1 + \phi t_{2j}\exp(\beta_2)}  = 0, \\
&\sum_{i=1}^{2} \sum_{j=1}^{n_{i}}\left\{-\sum_{l=0}^{Y_{ij} - 1} \left( \frac{1}{\phi + l \phi^2} \right) + \frac{Y_{ij}}{\phi} + \frac{\log(1 + \phi t_{ij} \exp(\beta_{i})) }{\phi^2} \right.\\
& \left.- \frac{(Y_{ij}\phi + 1)t_{ij} \exp(\beta_{i}) }{\phi + \phi^2 t_{ij}\exp(\beta_{i})} \right\}  =0.
\end{align*}
The solutions to those equations can be calculated using the multivariate Newton-Raphson algorithm.
The maximum likelihood estimator of the rate $\mu_i$ is then given by $\hat{\mu}_{i}=\exp(\hat{\beta}_{i})$.

\section*{Acknowledgement}
Tobias M{\"u}tze is supported by the DZHK (German Centre for Cardiovascular Research) under grant GOE SI 2 UMG Information and Data Management.
Tobias M{\"u}tze is grateful to Novartis Pharma AG (Basel) for support in form of an internship.


\begin{thebibliography}{00}

\bibitem[Shoben and Emerson(2014)]{shoben2014violations}
Shoben~AB and Emerson~SS.
Violations of the independent increment assumption when using generalized estimating equation in longitudinal group sequential trials.
\textit{Statistics in Medicine} 2014; \textbf{33}: 5041--5056.

\bibitem[Jennison and Turnbull(2000)]{jennison2000group}
Jennison~C and Turnbull~BW (2000) Group sequential designs with applications to clinical trials.
Chapman \& Hall/CRC; 2000.

\bibitem[Whitehead(1997)]{whitehead1997design}
Whitehead~J (1997) The design and analysis of sequential clinical trials.
John Wiley \& Sons; 1997.

\bibitem[Wassmer and Brannath(2016)]{wassmer2016group}
Wassmer~G and Brannath~W (2016) Group sequential and confirmatory adaptive designs in clinical trials.
Springer; 2016.

\bibitem[Sormani et al.(1999)]{sormani1999modelling}
Sormani~MP, Bruzzi~P, Miller~DH, Gasperini~C, Barkhof~F, and Filippi~M.
Modelling MRI enhancing lesion counts in multiple sclerosis using a negative binomial model: implications for clinical trials.
\textit{Journal of the Neurological Sciences} 1999; \textbf{163}: 74--80.

\bibitem[Keene et al.(2007)]{keene2007analysis}
Keene~ON, Jones~MRK, Lane~PW, and Anderson~J.
Analysis of exacerbation rates in asthma and chronic obstructive pulmonary disease: example from the TRISTAN study.
\textit{Pharmaceutical Statistics} 2007; \textbf{6}: 89--97.

\bibitem[Keene et al.(2008)]{keene2008statistical}
Keene~ON, Calverley~PMA, Jones~PW, Vestbo~J, and Anderson~JA.
Statistical analysis of exacerbation rates in COPD: TRISTAN and ISOLDE revisited.
\textit{European Respiratory Journal} 2008; \textbf{32}: 17--24.

\bibitem[Keene et al.(2009)]{keene2009methods}
Keene~ON, Vestbo~J, Anderson~J, Calverley~PMA, Celli~B, Ferguson~GT, Jenkins~C, and Jones~PW.
Methods for therapeutic trials in COPD: lessons from the TORCH trial.
\textit{European Respiratory Journal} 2009; \textbf{34}: 1018--1023.

\bibitem[Rod{\'e}s et al.(2015)]{rodes2015effect}
Rod{\'e}s-Cabau~J, Horlick~E, Ibrahim~R, Cheema~AN, Labinaz~M, Nadeem~N, Osten~M, C{\^o}t{\'e}~M, Marsal~JR, Rivest~D, Marrero~A, Houde~C.
Effect of clopidogrel and aspirin vs aspirin alone on migraine headaches after transcatheter atrial septal defect closure: the CANOA randomized clinical trial.
\textit{The Journal of the American Medical Association} 2015; \textbf{314}: 2147--2154.

\bibitem[Schlesinger et al.(2011)]{schlesinger2011canakinumab}
Schlesinger~N, Mysler~E, Lin~HY, De Meulemeester~M, Rovensky~J, Arulmani~U, Balfour~A, Krammer~G, Sallstig~P, So~A.
Canakinumab reduces the risk of acute gouty arthritis flares during initiation of allopurinol treatment: results of a double-blind, randomised study.
\textit{Annals of the Rheumatic Diseases} 2011; \textbf{70}: 1264--1271.

\bibitem[Rogers et al.(2014)]{rogers2014analysing}
Rogers~JK, Pocock~SJ, McMurray~JJV, Granger~CB, Michelson~EL, {\"O}stergren~J, Pfeffer~MA, Solomon~SD, Swedberg~K, and Yusuf~S..
Analysing recurrent hospitalizations in heart failure: a review of statistical methodology, with application to CHARM-Preserved.
\textit{European Journal of Heart Failure} 2014; \textbf{16}: 33--40.

\bibitem[Cook and Lawless(1996)]{cook1996interim}
Cook~RJ and Lawless~JF.
Interim monitoring of longitudinal comparative studies with recurrent event responses.
\textit{Biometrics} 1996; \textbf{52}: 1311--1323.

\bibitem[Cook and Farewell(1996)]{cook1996incorporating}
Cook~RJ and Farewell~VT.
Incorporating surrogate endpoints into group sequential trials.
\textit{Biometrical Journal} 1996; \textbf{38}: 119--130.

\bibitem[Jiang(1999)]{jiang1999group}
Jiang~W.
Group sequential procedures for repeated events data with frailty.
\textit{Journal of Biopharmaceutical Statistics} 1999; \textbf{9}: 379--399.

\bibitem[Xia and Hoover(2007)]{xia2007procedure}
Xia~Q and Hoover~DR.
A procedure for group sequential comparative Poisson trials.
\textit{Journal of Biopharmaceutical Statistics} 2007; \textbf{16}: 869--881.

\bibitem[Cook et al.(2010)]{cook2010sequential}
Cook~RJ, Grace~YY, and Lee~KA.
Sequential Testing with Recurrent Events over Multiple Treatment Periods.
\textit{Statistics in Biosciences} 2010; \textbf{2}: 137--153.

\bibitem[Ponikowski et al.(2016)]{Ponikowski2016esc}
Ponikowski~P, Voors~AA, Anker~SD, and others.
2016 ESC Guidelines for the diagnosis and treatment of acute and chronic heart failure.
\textit{European Journal of Heart Failure} 2016; \textbf{18}: 891-975.

\bibitem[CHMP(2016)]{CHMP2016esc}
Committee for Medicinal Products for Human Use (CHMP).
Guideline on clinical investigation of medicinal products for the treatment of chronic heart failure.
\url{https://goo.gl/HHgONb} Accessed: 2016/10/20.

\bibitem[Rogers et al.(2012)]{rogers2012eplerenone}
Rogers~JK, McMurray~JJV, Pocock~SJ, Zannad~F, Krum~H, van Veldhuisen~DJ, Swedberg~K, Shi~H, Vincent~J, and Pitt~B.
Eplerenone in patients with systolic heart failure and mild symptoms analysis of repeat hospitalizations.
\textit{Circulation} 2012; \textbf{126}: 2317--2323.

\bibitem[Rogers et al.(2014)]{rogers2014effect}
Rogers~JK, Jhund~PS, Perez~AC, B{\"o}hm~M, Cleland~JG, Gullestad~L, Kjekshus~J, van Veldhuisen~DJ, Wikstrand~J, Wedel~H, and others.
Effect of rosuvastatin on repeat heart failure hospitalizations: the CORONA Trial (Controlled Rosuvastatin Multinational Trial in Heart Failure).
\textit{JACC: Heart Failure} 2014; \textbf{2}: 289--297.

\bibitem[Udelson(2011)]{udelson2011}
Udelson~JE.
Heart Failure with Preserved Ejection Fraction.
\textit{Circulation} 2011; \textbf{124}: 540--543.

\bibitem[Redfield(2016)]{redfield2016}
Redfield~MM.
Heart Failure with Preserved Ejection Fraction.
\textit{New England Journal of Medicine} 2016; \textbf{375}: 1868--1877.

\bibitem[Yusuf et al.(2003)]{yusuf2003effects}
Yusuf~S, Pfeffer~MA, Swedberg~K, Granger~CB, Held~P, McMurray~JJV, Michelson~EL, Olofsson~B, {\"O}stergren~J, and others.
Effects of candesartan in patients with chronic heart failure and preserved left-ventricular ejection fraction: the CHARM-Preserved Trial.
\textit{The Lancet} 2003; \textbf{362}: 777--781.

\bibitem[Mehta et al.(2009)]{mehta2009optimizing}
Mehta~C, Gao~P, Bhatt~DL, Harrington~RA, Skerjanec~S, and Ware~JH.
Optimizing trial design: sequential, adaptive, and enrichment strategies.
\textit{Circulation} 2009; \textbf{119}: 597--605.

\bibitem[Jackson et al.(2016)]{jackson2015improving}
Jackson~N, Atar~D, Borentain~M, Breithardt~G, van Eickels~M, Endres~M, Fraass~U, Friede~T, Hannachi~H, Janmohamed~S and others.
Improving clinical trials for cardiovascular diseases: A position paper from the Cardiovascular Round Table of the European Society of Cardiology.
\textit{European Heart Journal} 2016; \textbf{37}: 747--754.

\bibitem[Filippatos et al.(2017)]{filippatos2017independent}
Filippatos~GS, de Graeff~P, Bax~JJ, Borg~JJ, Cleland~JGF, Dargie~HJ, Flather~M, Ford~I, Friede~T, Greenberg~B and others.
Independent academic Data Monitoring Committees for clinical trials in cardiovascular and cardiometabolic diseases.
\textit{European Journal of Heart Failure} 2017; \textbf{19}: 449-456.

\bibitem[Cohen and Rudick(2007)]{cohen2007multiple}
Cohen~JA, Rudick~RA.
Multiple sclerosis therapeutics.
\textit{CRC Press} 2007.

\bibitem[Selmaj et al.(2013)]{selmaj2013siponimod}
Selmaj~K, Li~SKB, Hartung~HP, Hemmer~B, Kappos~L,  Freedman~MS, St{\"u}ve~O, Rieckmann~P, Montalban~X, Ziemssen~T.
Siponimod for patients with relapsing-remitting multiple sclerosis (BOLD): an adaptive, dose-ranging, randomised, phase 2 study.
\textit{The Lancet Neurology} 2013; \textbf{12}: 756--767.

\bibitem[Cook and Lawless(2007)]{cook2007statistical}
Cook~RJ and Lawless~J.
The statistical analysis of recurrent events.
\textit{Springer Science \& Business Media} 2007.

\bibitem[Lawless(1987)]{lawless1987negative}
Lawless~JF.
Negative binomial and mixed Poisson regression.
\textit{Canadian Journal of Statistics} 1987; \textbf{15}: 209--225.

\bibitem[Gallo et al.(2014)]{gallo2014alternative}
Gallo~P, Mao~L, and Shih~VH.
Alternative views on setting clinical trial futility criteria.
\textit{Journal of Biopharmaceutical Statistics} 2014; \textbf{24}: 976--993.

\bibitem[Scharfstein et al.(1997)]{scharfstein1997semiparametric}
Scharfstein~DO, Tsiatis~AA, and Robins~JM.
Semiparametric efficiency and its implication on the design and analysis of group-sequential studies.
\textit{Journal of the American Statistical Association} 1997; \textbf{92}: 1342--1350.

\bibitem[Jennison and Turnbull(1997)]{jennison1997group}
Jennison~C and Turnbull~BW.
Group-sequential analysis incorporating covariate information.
\textit{Journal of the American Statistical Association} 1997; \textbf{92}: 1330--1341.

\bibitem[Lan and DeMets(1983)]{lan1983discrete}
Lan~KKG and DeMets~DL.
Discrete sequential boundaries for clinical trials.
\textit{Biometrika} 1983; \textbf{70}: 659--663.

\bibitem[Proschan et al.(1992)]{proschan1992effects}
Proschan~MA, Follmann~DA, and Waclawiw~MA.
Effects of assumption violations on type I error rate in group sequential monitoring.
\textit{Biometrics} 1992; \textbf{48}: 1131--1143.

\bibitem[Aban et al.(2009)]{aban2009inferences}
Aban~IB, Cutter~GR, and Mavinga~N.
Inferences and power analysis concerning two negative binomial distributions with an application to MRI lesion counts data.
\textit{Computational Statistics \& Data Analysis} 2009; \textbf{53}: 820--833.

\bibitem[Farrington and Manning(1990)]{farrington1990test}
Farrington~CP and Manning~G.
Test statistics and sample size formulae for comparative binomial trials with null hypothesis of non-zero risk difference or non-unity relative risk.
\textit{Statistics in Medicine} 1990; \textbf{9}:1447--1454.

\bibitem[Tang et al.(2006)]{tang2006sample}
Tang~NS, Tang~ML, and Wang~SF.
Sample size determination for matched-pair equivalence trials using rate ratio.
\textit{Biostatistics} 2006; \textbf{8}:625--631.

\bibitem[M{\"u}tze et al.(2016)]{mutze2016design}
M{\"u}tze~T, Munk~A, and Friede~T.
Design and analysis of three-arm trials with negative binomially distributed endpoints.
\textit{Statistics in Medicine} 2016; \textbf{35}:505--521.

\bibitem[Pocock(1977)]{pocock1977group}
Pocock~SJ.
Group sequential methods in the design and analysis of clinical trials.
\textit{Biometrika} 1977; \textbf{64}:191--199.

\bibitem[M{\"u}tze et al.(2017)]{mutze2017studentized}
M{\"u}tze~T, Konietschke~F, Munk~A, and Friede~T.
A studentized permutation test for three-arm trials in the `gold standard' design.
\textit{Statistics in Medicine} 2017; \textbf{36}:883--898.

\bibitem[Placzek and Friede(2017)]{placzek2017clinical}
Placzek~M and Friede~T.
Clinical trials with nested subgroups: Analysis, sample size determination and internal pilot.
\textit{Statistical Methods in Medical Research} 2017 \url{https://doi.org/10.1177/0962280217696116}.

\bibitem[Spiessens et al.(2010)]{spiessens2010adjusted}
Spiessens~B and Debois~M.
Adjusted significance levels for subgroup analyses in clinical trials.
\textit{Contemporary Clinical Trials} 2010; \textbf{31}:647--656.

\bibitem[Graf et al.(2017)]{graf2017robustness}
Graf~AC, Wassmer~G, Friede~T, Gera~RG, and Posch~M.
Robustness of testing procedures for confirmatory subpopulation analyses based on a continuous biomarker.
\textit{Submitted to Statistical Methods in Medical Research} 2017.

\bibitem[Jennison (1992)]{jennison1992}
Jennison~C.
Bootstrap tests and confidence intervals for a hazard ration when the number of observed failures is small, with applications to group sequential survival studies.
\textit{Computing Science and Statistics: Proceedings of the 22nd Interface Conference} 1992; \textbf{1}:89--97.

\bibitem{github2017gscounts}
M{\"u}tze~T.
R package gscounts. 2017.
\url{https://github.com/tobiasmuetze/gscounts}

\bibitem[Kieser and Hauschke(2005)]{kieser2005assessment}
Kieser~M, Hauschke~D.
Assessment of clinical relevance by considering point estimates and associated confidence intervals.
\textit{Pharmaceutical Statistics} 2005; \textbf{4}:101--107.

\bibitem[Fisch et al.(2015)]{fisch2015bayesian}
Fisch~R, Jones~I, Jones~J, Kerman~J, Rosenkranz~GK, and Schmidli~H.
Bayesian design of proof-of-concept trials.
\textit{Therapeutic Innovation \& Regulatory Science} 2015; \textbf{49}:155--162.

\bibitem[Gsponer et al.(2014)]{gsponer2014practical}
Gsponer~T, Gerber~F, Bornkamp~B, Ohlssen~D, Vandemeulebroecke~M, Schmidli~H.
A practical guide to Bayesian group sequential designs.
\textit{Pharmaceutical Statistics} 2014; \textbf{13}:71--80.

\bibitem[Asendorf et al.(2017)]{asendorf2017}
Asendorf~T, Henderson~R, Schmidli~H, and Friede~T.
Modelling and sample size reestimation for longitudinal count data with incomplete follow up.
\textit{(Submitted for publication)}.

\bibitem{CHMPcovariates}
{Committee for Medicinal Products for Human Use (CHMP)}.
Guideline on adjustment for baseline covariates in clinical trials.
\url{https://goo.gl/YIJHEh} (accessed November 20, 2016).

\bibitem[McCullagh and Nelder(1989)]{mccullagh1989generalized}
McCullagh~P and Nelder~JA.
Generalized Linear Models.
\textit{Chapman \& Hall} 1989.

\bibitem[Scharfstein and Tsiatis(1998)]{scharfstein1998use}
Scharfstein~DO and Tsiatis~AA.
The use of simulation and bootstrap in information-based group sequential studies.
\textit{Statistics in Medicine} 1998; \textbf{17}: 75--87.

\bibitem[Mehta and Tsiatis(2001)]{mehta2001flexible}
Mehta~CR and Tsiatis~AA.
Flexible sample size considerations using information-based interim monitoring.
\textit{Drug Information Journal} 2001; \textbf{35}: 1095--1112.

\bibitem[Parmar et al.(2001)]{parmar2001monitoring}
Parmar~MKB, Griffiths~GO, Spiegelhalter~DJ, Souhami~RL, Altman~DG, and van der Scheuren~E.
Monitoring of large randomised clinical trials: a new approach with Bayesian methods.
\textit{The Lancet} 2001; \textbf{358}:375--381.

\bibitem[Friede and Schmidli(2010)]{friede2010blindeda}
Friede~T and Schmidli~H.
Blinded sample size reestimation with count data: methods and applications in multiple sclerosis.
\textit{Statistics in Medicine} 2010; \textbf{29}:1145--1156.

\bibitem[Friede and Schmidli(2010)]{friede2010blindedb}
Friede~T and Schmidli~H.
Blinded sample size reestimation with negative binomial counts in superiority and non-inferiority trials.
\textit{Methods of Information in Medicine} 2010; \textbf{49}:618--624.

\end{thebibliography}
\end{document}